# Characterization of local deformation around hydrides in Zircaloy-4 using conventional and high angular resolution electron backscatter diffraction

Ruth M. Birch[1,2*], James O. Douglas[1], T. Ben Britton[1,2]

1. Department of Materials, Imperial College London, Exhibition Road, London, UK, SW7 2AZ
2. Department of Materials Engineering, University of British Columbia, Frank Forward Building, 309-6350 Stores Road, Vancouver, BC, Canada V6T 1Z4

*corresponding author: ruth.birch@ubc.ca

# 1  Abstract

Zircaloy-4 is used as a fuel cladding material for water reactors, as it has good mechanical properties, corrosion resistance, and a low thermal neutron absorption cross section. However, the mechanical performance of Zircaloy-4 can be reduced during service due to hydrogen uptake and hydride formation. These hydrides are brittle, and often reduce the strength and toughness of materials as well as increase susceptibility to delayed hydride cracking (DHC). In this work, large grain Zircaloy-4 with hydrides was prepared and then cross sectioned using cryo-ion beam polishing, using plasma focused ion beam (pFIB) and broad ion beam (BIB) approaches to enable the preparation of a very high quality flat surface with no preferential etching of either the hydride or zirconium metal (typically metallographic polishing preferentially removes hydrides). Conventional and high angular resolution electron backscatter diffraction (EBSD) analysis were then used to explore morphology, deformation fields, and orientation relationships between the zirconium matrix and hydrides. Four maps were collected for analysis which included hydrides near grain boundaries: (a) where the hydride smoothly decorates across two of the connecting boundaries near a triple junction; (b) where the hydride smoothly decorates the boundary; (c) a mixture of smooth decoration of the interface and protrusion into the grains; (d) fine scale hydride that protrudes into one grain. This work highlights that incompatibility of the hydride within the zirconium matrix is strongly linked to the orientation relationship of the hydride and matrix, and the grain boundary character. These results may enable enhanced understanding of the role of hydrides in fracture as well as stress-induced hydride reorientation and DHC susceptibility.

# 2  Keywords

Strain measurement; stress fields; electron microscopy; nuclear materials

# 3 Introduction

Zirconium alloys are used extensively in the nuclear industry for fuel rods and other structural components as a structural metal that does not absorb neutrons from the reactor, withstands corrosion at high temperatures for long periods, and maintains its structural integrity under intense radiation [4]. Despite their continued use over the past 70 years [4], hydrogen ingress in-service remains a challenge to improve fuel efficiency and lifetime management of nuclear fuel systems. Hydride ingress can result in the formation of zirconium hydride which can degrade fuel cladding performance [5].

For zirconium alloys, the density and location of hydrides that form is affected by hydrogen content, temperature gradients, stress and cooling rate [6], [7]. During reactor operation, the fuel can vary in temperature (e.g. during operation, and when powering up or powering down the reactor) which creates thermal cycling. In the heating cycle, existing hydride precipitates can dissolve into the zirconium matrix and then the hydrogen can diffuse throughout the material. In the cooling cycle, new hydrides may reprecipitate in the matrix. Overall, this process results in the 'movement of hydrides' through the material, as controlled by the diffusion of hydrogen through the material (e.g. from hot to cold regions) and the driving forces for precipitation of new hydrides. One of these driving forces for hydride growth is the stress state of the matrix, and this can result in a process known as 'hydride reorientation' where hydrides reorient to align with their long axis perpendicular to the principal stress after a thermal cycle. This is particularly problematic as it can provide an easy crack path through the material (due to easy crack initiation or fast propagation along the hydride/metal interface or within a hydride bundle) and this has been a focus of study when addressing delayed hydride cracking (DHC).

Hydrides form as a solid phase precipitating reaction when there is excess hydrogen solute in the α-zirconium phase. At room temperature, and with sufficiently slow cooling, typically δ-phase hydride plates will form, and the cooling rate and grain size can affect the size and location of the nucleation of hydrides [2]. This solid-state phase transform typically results in the nucleation of hydrides with a specific orientation relationship (OR) with the parent α-zirconium phase, as described in Table 1.

The hydrides have a larger lattice parameter than the α-zirconium phase and this creates local misfit. Carpenter [8] calculated the theoretical (stress-free) misfit strain for the hydride based on the orientation relationship [9] with the Zr matrix, as detailed in Table 1.

A stress (or lattice strain) reorientation of the hydride precipitates has been described in the literature and correlated with the associated local lattice misfit strain. Carpenter [8] suggests that this local strain can be relaxed by the generation of dislocation loops near the hydride tips. Experiments by Bailey [10] show that dislocation loops are present around γ phase hydride needles and the number of loops present in the work match well with the number required to accommodate the strain around the hydride precipitate.

The observation of hydride reorientation in the presence of a macroscopic applied stress, as well as alignment of hydrides ahead of a crack tip associated with a critical stress intensity [11] implies that the stress (and/or strain) field of the hydride/matrix could assist in understanding the propensity of hydrides to reorientate and precipitate in specific locations within a microstructure.

Table 1 - Dilatational misfit values for δ hydrides based on phase and direction in the Zr matrix. Compiled from [8].

| Hydride phase | Orientation relationship | Direction (in Zr) | Misfit[1] (%) |
|---|---|---|---|
| Delta (δ) | Multiple ORs – common close packed direction between: $[1\bar{1}0]_\delta \parallel [11\bar{2}0]_\alpha$ | $[0001]$ | 7.20 |
| | | $[11\bar{2}0]$ | 4.58 |
| | | $[1\bar{1}00]$ | 4.58 |

To date, the literature contains reports of hydride formation that focus on small lenticular hydrides within grains and hydrides within the bulk Zr matrix, where different experimental techniques have been used to measure the strain field around hydrides. For example, Barrow et al. [12] evaluated the interfacial strain around hydrides in Zircaloy-2 using nano-beam electron diffraction (NBED) and electron energy loss spectroscopy (EELS). Kerr et al. [13] used *in situ* synchrotron X-ray diffraction combined with mechanical testing to enable analysis of the lattice strain evolution under an applied load (compressive and tensile). Allen et al. [14] used high energy X-ray diffraction (HRXRD) to investigate strains related to hydrides formed at a notch (stress concentration) in Zr-2.5Nb pressure tube material.

Modelling of the stress field has been carried out by Kerr et al. [14], who use a finite element model for the stress state and the effect of hydrides on the strain profile at the notch tip; and Reali et al.[15], who developed a discrete dislocation plasticity (DDP) model to look at the stress at the hydride/matrix interface and investigate plastic relaxation.

Whilst most of the experimental work has assumed predominantly δ phase hydrides as being present, a potential source of stress variations around the small lenticular hydrides is a phase change at the hydride tips, from δ phase hydride to gamma (γ) phase hydride. γ phase hydrides have a different morphology and would result in slightly different misfit strain values (see [8]). Barrow et al. [12] show observation of a phase change at the hydride tips using EELS and suggest that this is due to local hydrogen enrichment at the matrix–hydride interface as a result of tensile interfacial strains. However, whilst there may be some γ phase present alongside δ phase ([16], [17]), it has also been suggested that at higher hydrogen content, hydrides grow out as delta phase [18]. The stability of gamma phase hydride has been a long standing topic of discussion, with recent observations [19] suggesting that gamma phase hydride is only stable at low temperatures. In this work, we focus on the δ phase as those are the hydrides observed via electron backscatter diffraction (EBSD) for the chosen precipitation route (see [2]).

---

[1] Misfit = volume change relative to the untransformed Zr.

Generally, from the combination of experiments and simulations within the literature we see that the misfit of the hydrides causes significant strain around the lenticular hydrides and how that strain is accommodated depends on the hydride phase. Furthermore, the strain field is strain is heterogenous. These observations imply that the superposition of strain fields from adjacent hydrides may result in more complicated resultant strain fields, and affect nucleation and growth of hydrides, as well as further complications when the hydrides form within neighbouring grains or at grain boundaries. This strain field likely impacts the build-up of lattice strain in the matrix, in turn affecting further nucleation and growth (or dissolution) of hydrides in the matrix, as well as clustering of hydrides to form hydride packets. These phenomena likely control hydride reorientation and delayed hydride cracking.

From this literature review, the magnitude and spatial variation of the misfit (elastic) strain of larger hydride packet structures and near interfaces such as grain boundaries is not well understood, which limits our understanding of the micro- to macro-scale hydride reorientation mechanisms and this motivates the present work, where we use conventional and high angular resolution electron backscatter diffraction (EBSD and HR-EBSD) to evaluate the local elastic strain (i.e. gradients of lattice rotation and deviatoric stress) around δ phase hydrides located at and near grain boundaries in Zircaloy-4.

## 4 Methods

Two polycrystalline Zircaloy-4 samples were cut from a recrystallised plate, which had a grain size of ~11 µm. The samples were heat treated to grow large 'blocky-α' grains by annealing for 2 weeks at 800 °C and then air cooled (see [1] for more details). Each sample was individually electrolytically hydrided in 1.5 wt% $H_2SO_4$ using a current density of 2 kA/$m^2$ at 65 °C for 24 h (see [2] for the method and a schematic). To control the average hydrogen concentration within each sample, charging was only performed on a limited surface area of each sample using a Lacomit varnish. After electrolytic hydriding, hydrogen was diffused into the bulk samples by annealing at a temperature within the α-Zr phase field, i.e. above the hydride dissolution temperature and below the beta transus temperature. After annealing, the samples were cooled at a rate to promote the formation of δ phase hydrides, mainly on the grain boundaries (as determined from prior work [2]). Sample A was hydrided on one side, and contained a predicted hydrogen content around 130 wt-ppm, based on prior work [2] whilst sample B was hydrided on two sides and expected to have a higher hydrogen content – see the Supplementary Information for more details.

Next, an initial metallographic polishing process was performed. Each sample was mechanically ground using SiC papers with 800, 1200 & 2000 grit size, followed by mechanical polishing for up to 4.5 hrs using an MD-Chem polishing pad and an OPS based polishing solution mixed at ratio of 5 OPS : 6 $H_2O$ : 1 $H_2O_2$.

This metallographic polishing step does not polish the hydride and matrix equally, and often leaves significant surface topography which makes EBSD difficult (as one phase has poor quality patterns, and the roughness causes shadowing of the patterns). Therefore, for EBSD analysis, final polishing was performed using ion beam polishing. This was carried out at cryogenic temperatures to reduce ion-surface damage and secondary formation of hydrides (a known issue for Zr- and Ti-alloys, see [20], [21]).

For the present work, two ion beam polishing methods were employed:

- Sample A was prepared using plasma focussed ion beam polishing (pFIB) using the the recipe detailed in Table 2. For polishing at under cryo-conditions, the section of Zircalloy was mechanically clamped using a copper clip typically used for atom probe tomography silicon coupon samples (Cameca), with silver paint used to give additional mechanical support. Helios Hydra CX (5CX) plasma FIB from Thermofisher Scientific (TFS) with an Aquilos cryo stage, modified by TFS to accommodate both a conventional dove-tailed holder and commercial atom probe sample holder from Cameca. The stage was cooled to approximately 90 K by using a circulation of gaseous nitrogen passing through a heat exchanged system within a liquid nitrogen (LN2) Dewar. A $N_2$ gas flow of 180 mg/s was used to achieve the base temperature for sample preparation in this work.

*Table 2 - pFIB polishing recipe details for hydrided Zircaloy-4.*

| Step | | Aim | Details |
|---|---|---|---|
| 1 | Rough mill | Setup for polishing | - Room temperature<br>- 30 kV 2.5 µA Xe for rough mill<br>- Deposit Pt at 12 kV at 60 nA for 5 µm thick |
| 2 | Initial mill | Find ROI grain boundary region | - Mill at 30 kV Xe at 0.5 µA |
| 3 | Polish | Cryo-final polishing | - Cryo temperature<br>- Switch to Ar plasma<br>- Tilt to 53°<br>- 30 kV and 0.5 µA<br>- 12 kV and 200 nA for polishing<br>- 12 kV and 28 nA for polishing<br>- 12 kV and 14 nA for polishing |

- Sample B was prepared by broad ion beam polishing (BIB) using a Gatan PECSII machine polishing in a cross-section configuration. Building on prior work [22] the BIB recipe used was: 8 keV, 0 ° gun tilt, single modulation – 90 ° sector, rotation speed: 1 rpm, duration: 10 hrs, and cooled using a liquid nitrogen Dewar. Cross sectioning of the sample was performed by masking near the surface with titanium blade and polishing away the Zircaloy-4 sample to reveal a large area and high quality cross section near the metallographically polished surface (see Figure 1 for a schematic and example electron micrograph of a sample with a polished edge).

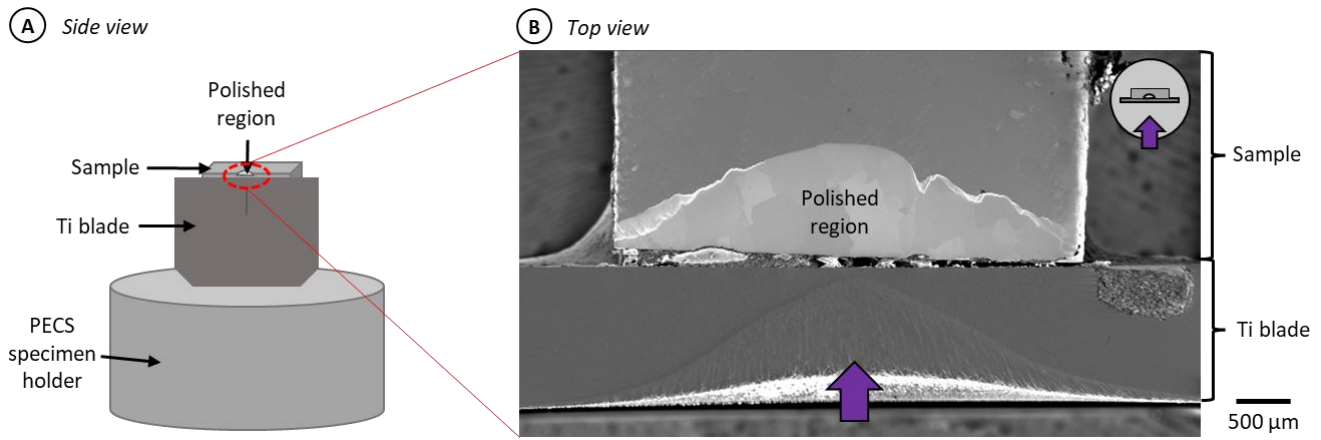

*Figure 1 - Cross section BIB polishing of large grain Zircaloy-4. A shows a schematic showing sample behind Ti blade with polished region indicated (side view) and B shows top view setup for polishing (SE image). The purple arrow indicates the ion beam polishing direction.*

For a comparison of the two polishing methods, the reader is directed towards the discussion in the supplementary material.

Electron backscatter diffraction (EBSD) data was collected using an FEI Quanta FEG 650 SEM equipped with a Bruker eFlashHD2 EBSD detector. EBSD maps were captured with patterns stored to disk at 2x2 binning (i.e. 800 x 600 pixels) using an estimated probe current of ~10 nA, and an exposure time of 150 ms. EBSD mapping was performed with a step size of 0.08 µm for sample A, and 0.10 µm for sample B. Online orientation analysis was performed using eSprit v2.2.

The EBSD data was initially analysed using MTEX [23] to measure the crystal orientations and determine the phase for each point. This was then used for polycrystalline high angular resolution EBSD (HR-EBSD) analysis using the XEBSD software code, developed by Britton and Wilkinson [24].

In brief, the XEBSD code follows the HR-EBSD method of comparing two diffraction patterns to measure changes in (deviatoric) elastic strain and lattice rotation between them (based upon the initial work described in [25] and [26]). High precision zone axis shift measurements using sub-region-based image correlation, and one pass of pattern remapping to remove artifacts associated with large image shifts. In this work, fifty 128 x 128 pixel regions of interest (ROI) were used, with one ROI at the EBSD image centre, nineteen in a ring around this, and the remainder randomly scattered across the pattern. From these pattern shifts (measured with sub-pixel precision) changes in lattice rotation and (deviatoric) lattice strain are calculated with a precision of $1\times10^{-4}$ rads and $1\times10^{-4}$ in strain. The HR-EBSD method cannot measure volumetric strain differences between the two patterns, and therefore separation of the normal strain tensor components is calculated using Hooke's law, by assuming that the out of plane normal stress is equal to zero and input (anisotropic) elastic constants based using the crystal orientation and a stiffness matrix using the method described in [27]. For the zirconium (hexagonal symmetry) the following single crystal stiffness constants (in GPa) were used, as taken from [28] : $C_{11}$ = 143.4, $C_{12}$ = 72.8, $C_{13}$ = 65.3 , $C_{33}$ = 164.8, $C_{55}$ = 35.3 . For the δ-phase zirconium hydride (cubic symmetry) the following single crystal stiffness values were used, as taken from [28]: $C_{11}$ = 81.0, $C_{12}$ = 197.0, $C_{44}$ = 56.0.

Note that for polycrystalline samples, XEBSD performs the analysis grain-by-grain and the relative lattice rotation and elastic strain state between grains (the so-called Type II strains) are not known. This means that maps of relative lattice rotation, strain and stress are presented later with respect to the reference point within each region.

The HR-EBSD method produces two quality metrics that can be used to determined how successful the cross-correlation based analysis has been performed:

- The peak height (PH) is the geometric mean of the maximum value of cross correlation between each reference ROI and the remapped test ROI and within each grain this peak height is normalised to 1 for autocorrelation. A low value of peak height (typically less than 0.3) indicates poor matching and the result is likely to be incorrect.
- The mean angular error (MAE) is the (angular) difference between a forward calculation of the shifts between test and reference as expected from the (overdetermined) solution to strain and rotation between test and rotation, and the shifts measured at each ROI. A MAE value greater than the reported strain or lattice rotation tensor component indicates that the model used for matching shifts does not represent the shift vectors as measured. A high MAE can be often found when the reference and (remapped) test pattern do not correlate well, and can be caused due to pattern overlap and/or poor quality patterns, e.g. due to very large strain gradients within the interaction volume.

GND density is analysed using the Nye tensor approach, where dislocations consistent with the Zr-HCP deformation modes are used as potential dislocation types. The GND density is calculated using a L1 minimization based upon the line energy of dislocations that relate to the lattice curvature (the 6 spatial gradients of measured lattice rotation), assuming that the curvature can be supported by <a> screw, <a> basal, <a> prism, <a> pyramidal, <c+a> screw, and <c+a> pyramidal slip systems. For more information on this calculation, see ref [29].

Maps of GND density map tell us about the stored dislocation content associated with from significant incompatibility (e.g. from misfit strain), which we assume is accommodated by the formation and storage of (geometrically necessary) dislocations. Here we refer to GNDs as the net dislocation content that gives rise to a lattice curvature, and the density of statistically stored dislocation (SSDs) is not measured.

## 5   Results

Hydrides located at and near grain boundaries (grain boundary hydride) from four selected regions were investigated in detail. Due to the difference in polished region size and quality 30 x 15 µm for sample A (pFIB) vs. 2.3 x 0.6 mm for sample B (BIB), three of the examples were selected from sample B, and one from sample A.

These examples are divided into two types – examples where the hydrides smoothly decorate the grain boundary, and examples where the hydrides protrude into the matrix, as it is expected that the stress fields may differ significantly between these cases. A summary of the regions of interest studied is found in Table 3 below.

The hydrides in this sample are delta hydrides which should have an ideal orientation relationship of $(0001)_\alpha \parallel \{1\bar{1}1\}_\delta$ and $\langle 11\bar{2}0 \rangle_\alpha \parallel \langle 1\bar{1}0 \rangle_\delta$ [9].

*Table 3 - Regions of interest and hydride type contained within*

| ROI | Final Polish | Hydride Type | Description |
|---|---|---|---|
| A1 | pFIB | Smooth | Triple junction – only some grain boundaries have hydride |
| B1 | BIB | Smooth | Two grains with fully covered grain boundary |
| B2 | BIB | Protruding | Pair of hydrides at grain boundary, smoothly decorating in the middle and protruding into the grains at the ends |
| B3 | BIB | Protruding | Grain boundary with only hydride protruding into the matrix of one of the adjacent grains |

**5.1  Smooth grain boundary hydride**

There are two maps taken from where the hydride is smoothly decorating α-Zr grain boundaries, with no ingress of the hydride extending away from the grain boundary and into the matrix.

**5.1.1  Example 1 (B1 – BIB polished): grain boundary hydride**

The grain boundary hydride map, shown in Figure 2, consists of two large α grains with δ phase hydride on the grain boundary between, covering the complete length of the grain boundary imaged. The hydride thickness varies from approx. 0.5 µm to 2 µm in width, with a tapering around a (likely) secondary phase particle (SPP), and a small amount of α-Zr (labelled $\alpha_5$) contained within the hydride region.

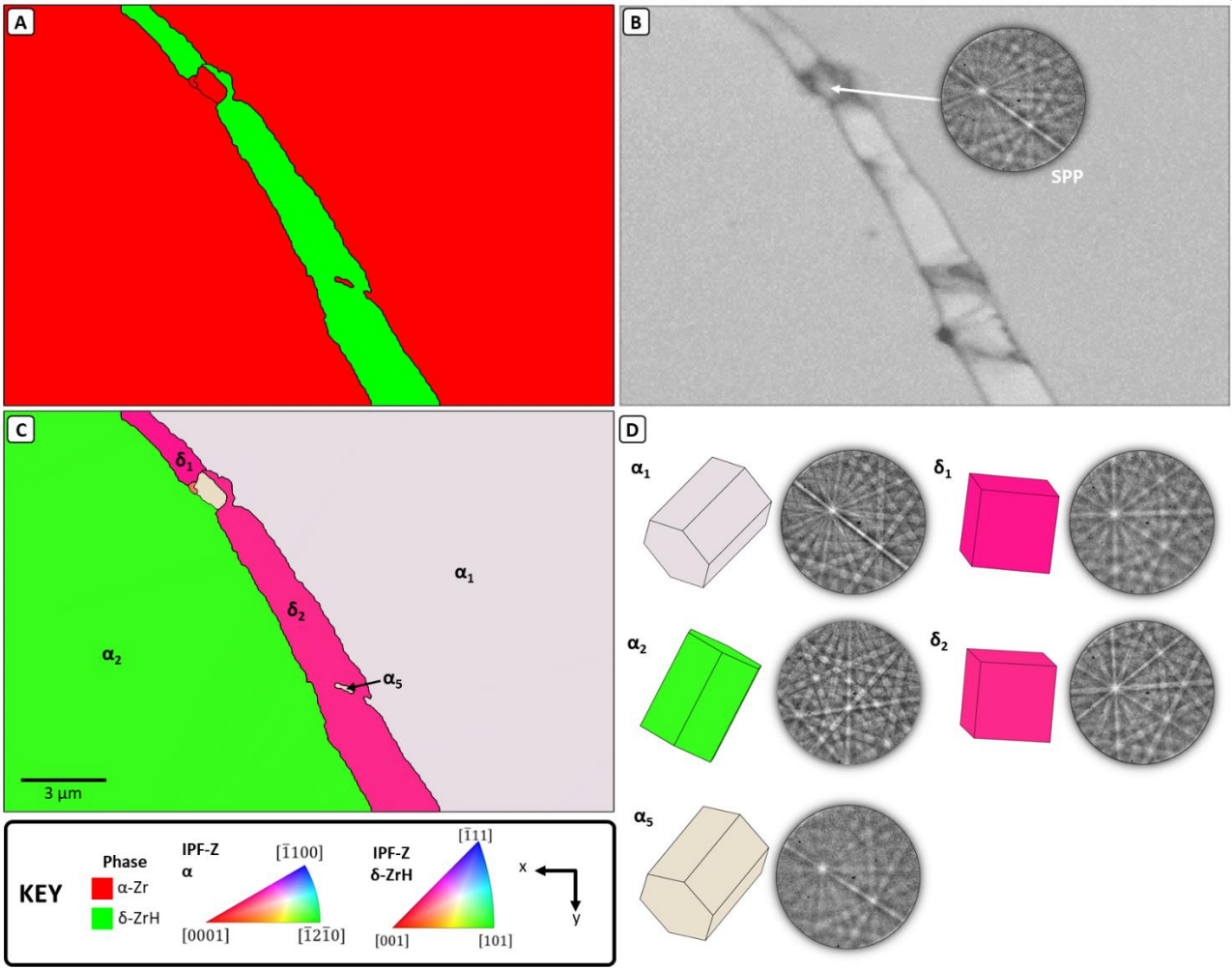

*Figure 2 – Orientation analysis of the grain boundary hydride map: (A) Indexed phase map; (B) pattern quality map (with probable SPP representative pattern); (C) crystal orientation (IPFZ coloured) along with labels for key grains; (D) prisms and representative patterns for the identified grains.* Figure 2 shows an overview of the orientations in the grain boundary hydride map. From the pattern quality (Figure 2 B) there is a small region that although indexed as α zirconium matrix, is likely to be an SPP, based on the pattern quality and orientation. The example EBSP for this region resembles that seen in Figure 2D for $α_1$ & $α_5$ which indicates that its orientation is related to the $α_1$ grain. This is likely a FeCr SPP with a hexagonal structure [18], [30].

Also seen in the pattern quality map (Figure 2B) are a number of darker lines within the hydride, indicating the presence of interfaces within the hydride, which is likely due to stacking of smaller hydride packets within the macro hydride [31].

The small orange grain in Figure 2C is not considered in this analysis as there is poor pattern quality at this point, which makes it difficult to tell if this is a grain and challenging to determine what phase it is.

For the α grains, $α_5$ has a very similar representative orientation to $α_1$ and is/was probably part of this grain prior to hydride formation (similar observations have been found by Wang et al. [9]).

In this cross section, the δ hydride is separated into two sections by the SPP, but each section (δ₁ and δ₂) has a similar orientation with respect to each other indicating that these hydride sections may be part of a larger hydride cluster in 3D, or there is an OR that relates the SPP and each hydride.

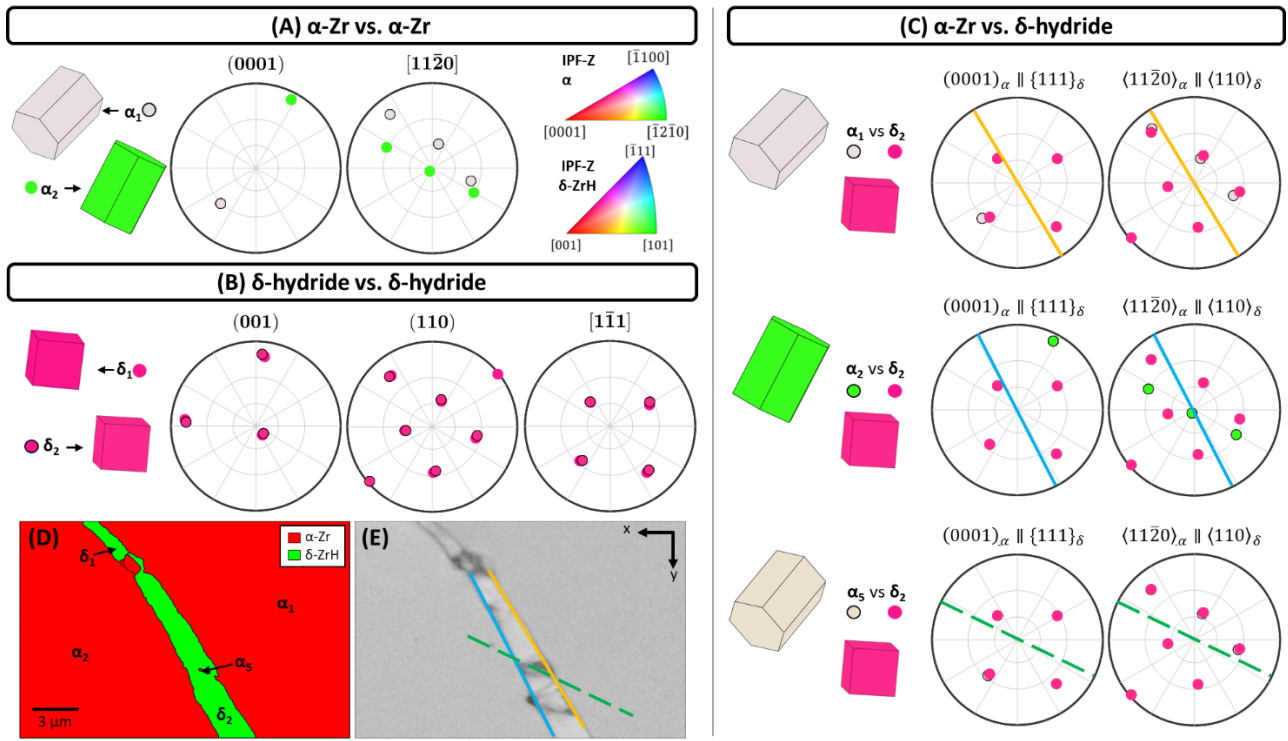

*Figure 3 – Orientation relationships in the grain boundary hydride map: Pole figure analysis of (A) α-Zr/α-Zr OR;(B) hydride/hydride OR; (C) α-Zr/δ-hydride OR with grain boundary traces overlaid; (D) phase map annotated with grain labels; (E) Grain boundary traces on pattern quality map.*

Figure 3 explores the orientation relationships between the α grains, δ-hydride grains and between the α-Zr and δ-hydride grains. There is no OR between $\alpha_1$ and $\alpha_2$, as shown in Figure 3 (A), and they have a <c> axis misalignment of 38 ° with the closest $\langle 11\bar{2}0 \rangle_\alpha$ directions 13 ° apart. Note that due to the similarity in orientation of $\alpha_1$ and $\alpha_5$, only $\alpha_1$ is used for the analysis in Figure 3 (A).

Figure 3 (B) shows that the two δ hydride grains are very close in orientation, and due to the poor pattern quality in the vicinity of the grain boundary (see Figure 2 (B)), these are likely part of the same hydride packet. As they are so similar, only $\delta_2$ is used for the α vs. δ analysis in Figure 3 (C). The orientations for $\alpha_1$ and $\delta_2$ match the hydride OR, whereas this OR is not met for $\alpha_2$ and $\delta_2$. This indicates that the hydride grew from $\alpha_1$. There is no discernible OR between $\alpha_2$ and the hydride. Interestingly, the α region within the hydride, $\alpha_5$, shows a near perfect match with the hydride OR.

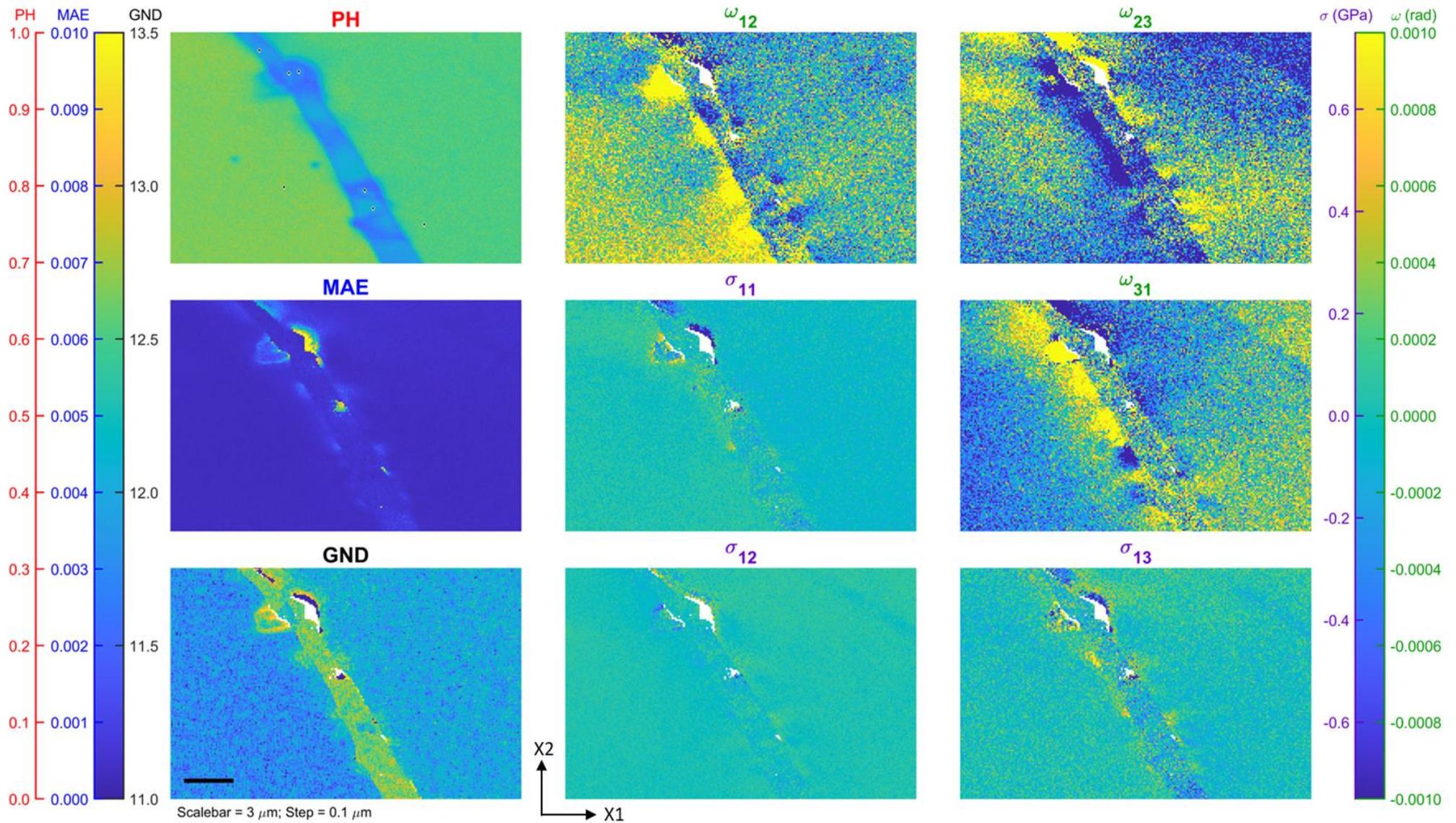

*Figure 4 HR-EBSD analysis of grain boundary hydride: PH = peak height with reference points per mapped region plotted; MAE = mean angular error; GND = sum of GND density across all slip systems, reported as $\log_{10}$(dislocation length/volume, in per $m^2$); components of the rotation ($\omega_{ij}$, in radians) and in-plane stress tensor ($\sigma_{ij}$). Maps are plotted against different colour scales. The stress tensor and rotation tensor maps are shown with respect to the reference points, as indicated on the PH map.*

Results from the HR-EBSD analysis are shown in Figure 4 for the grain boundary hydride, and they show the local variations in lattice rotation, in-plane stress and stored GND density. The peak height and MAE maps show that the SPP, hydride, and neighbour regions have poorer quality patterns and that the HR-EBSD analysis was less accurate for those regions. All the other maps have been filtered to only show reasonable values, using a filter to remove points with a PH < 0.3, and a MAE > $1 \times 10^{-3}$.

The rotation tensor maps $(\omega_{ij})$ show that there is a periodic variation of lattice rotation along the hydride metal interface, with notable positive and negative rotation components periodically along the hydride matrix interface. These fluctuations are seen long the interface between the hydride and $\alpha_1$, with a spacing of ~2 μm. Along the interface with $\alpha_2$, there is a more continuous variation in lattice rotation and the fluctuations along the interface are of smaller magnitude, except where the hydride has broken up (towards the bottom of the map) and associated with the SPP.

The stress field maps $(\sigma_{ij})$ show fluctuations of up to ~500 MPa at the interface, but these regions of high stress are limited to only ~1 μm from the interface and their spatial extent perpendicular to the interface is lower than the variations in lattice rotation.

In the GND map in Figure 4 there are regions of slightly higher GND density within the metal corresponding to the periodic fluctuations in stress and lattice rotation variation, as well as a higher GND density around the SPP particle.

### 5.1.2    Example 2: (A1 – pFIB polished): triple junction grain boundary hydride

The triple junction grain boundary map (Figure 5) contains a triple junction with grain boundary hydride decorating two of the branches. The maximum hydride width at the grain boundaries is ~1 μm. In this case, the hydride is more complicated, with four δ hydride grains in total, one between $\alpha_1$ and $\alpha_2$, and three along the grain boundary between $\alpha_2$ and $\alpha_3$.

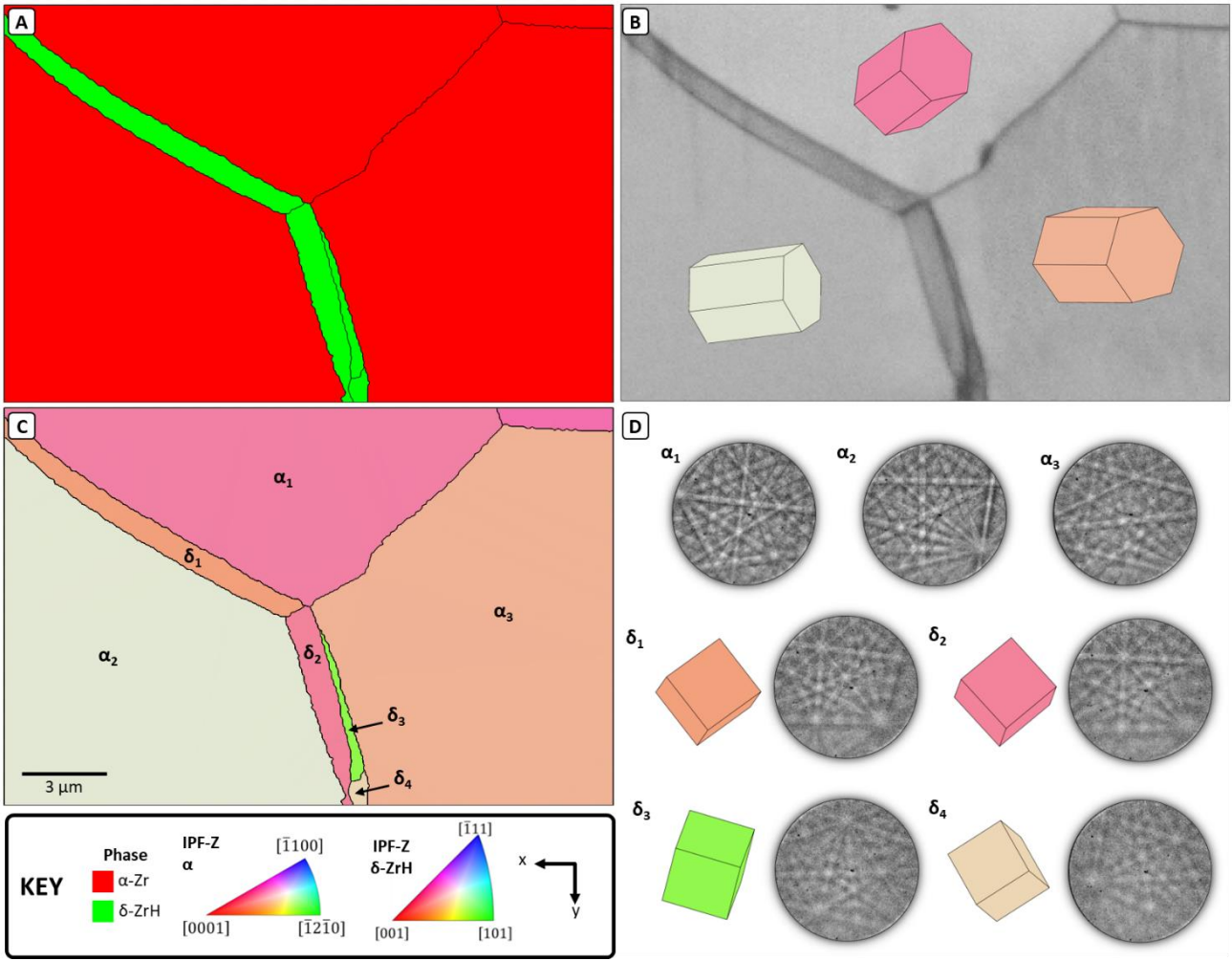

*Figure 5 - Orientation analysis of the triple junction grain boundary hydride map. (A) Indexed phase map; (B) pattern quality map (with probable SPP representative pattern); (C) crystal orientation (IPFZ coloured) along with labels for key grains; (D) prisms and representative patterns for the identified grains.*

The three α grains (shown in Figure 5) have relatively similar orientations, as indicated by the prisms and the IPF-Z colouring. Two of the δ hydrides ($δ_1$ and $δ_2$) have a very similar orientation as shown in Figure 5 D, whilst the remaining δ hydrides also appear related, with a rotation seen when comparing the two EBSPs.

Analysis of the α/α and α/δ orientations in this map are shown in Figure 6. It is interesting that the undecorated $α_1/α_3$ grain boundary has a very closely shared $[11\bar{2}0]_α$ direction, compared to the hydride decorated $α_1/α_2$ and $α_2/α_3$ grain boundaries where the α grains have a close, but not exact, shared $[11\bar{2}0]_α$ direction, as detailed in Table 4.

*Table 4 - Angles between α grain directions in the triple junction map – measured in degrees using MTEX.*

| α-grain pair | ∠ between $(0001)_α$ (°) | ∠ between $\langle 11\bar{2}0 \rangle_α$ (°) |
|---|---|---|
| $α_1$ and $α_2$ | 35.4 | 11.9 |
| $α_1$ and $α_3$ | 22.3 | 5.4 |
| $α_2$ and $α_3$ | 24.8 | 13.1 |

Analysis of the OR between the α grains and δ hydrides compared with the hydride OR (see Figure 6 B) indicates that $δ_1$ and $δ_2$ both originate from $α_2$, whilst $δ_3$ and $δ_4$ originate from $α_3$.

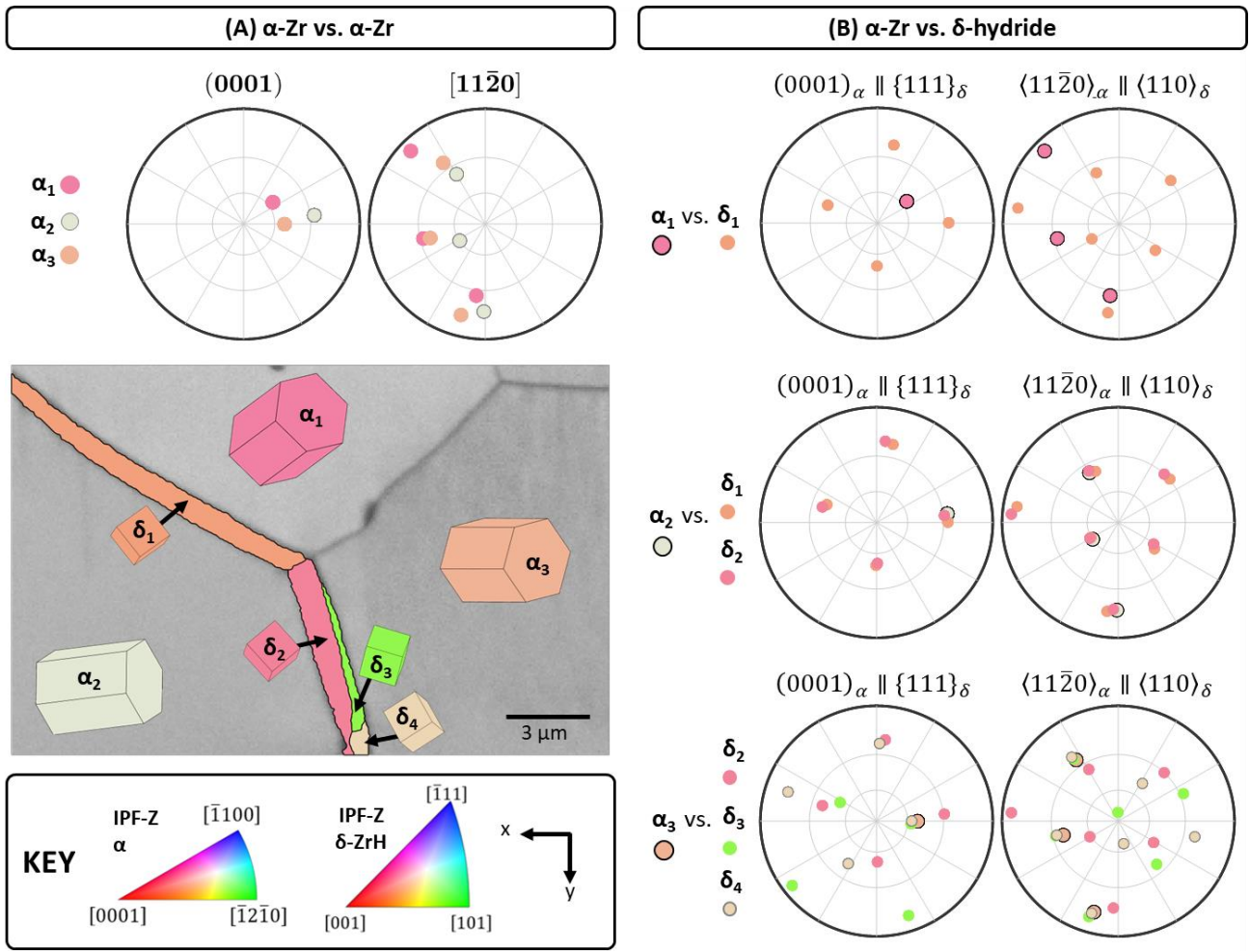

*Figure 6 - Orientation relationships in the triple junction grain boundary map: Pole figure analysis of (A) α-Zr/α-Zr OR; (B) α-Zr/δ-hydride ORs. Pattern quality map with hydride grains overlaid using IPFZ colouring and labelled prisms (IPFZ colouring) are also included for reference.*

It is possible also to characterise the misorientation between δ/δ hydride pairs, as shown in Figure 7. Here there are two cases: (1) between hydrides that originate from the same α grain ($δ_1$ & $δ_2$, $δ_3$ & $δ_4$); (2) between adjacent hydrides associated with different α grains (i.e. $δ_2$ and $δ_3$).

For the hydrides that originate from the same grain: $δ_1$ and $δ_2$ are closely aligned, with a misorientation of ~ 5 ° whilst $δ_3$ and $δ_4$ share a common $\langle 111 \rangle_δ$ direction, and multiple $\{110\}_δ$ planes, indicating that they are twins. It is likely that $δ_2$ grew from $δ_1$ first (hence the larger hydrides) and $δ_3$ grew from $δ_4$, with $δ_3$ being the accommodating hydride between $δ_2$ and $δ_4$, hence the tapering off. Looking at the trace of the grain boundary, and the perpendicular bisector, a $\langle 110 \rangle_δ$ direction is found on the bisector line for both δ in both cases. This indicates that the habit plane between these hydrides is a $\langle 110 \rangle_δ$ direction.

For the adjacent hydrides associated with different α grains, the only shared direction (of those evaluated) is $\langle 112 \rangle_δ$ where there is a 4 ° difference. Looking at the grain boundary trace and the perpendicular bisector, again the $\langle 112 \rangle_δ$ direction appears to be key, with $\langle 112 \rangle_δ$ directions from both $δ_2$ and $δ_3$ on the bisector line, indicating that this is the habit plane in this case. Note that the other similar pair, $δ_2$ & $δ_4$ pair is not analysed as the interface has a high degree of curvature preventing use of a grain boundary trace.

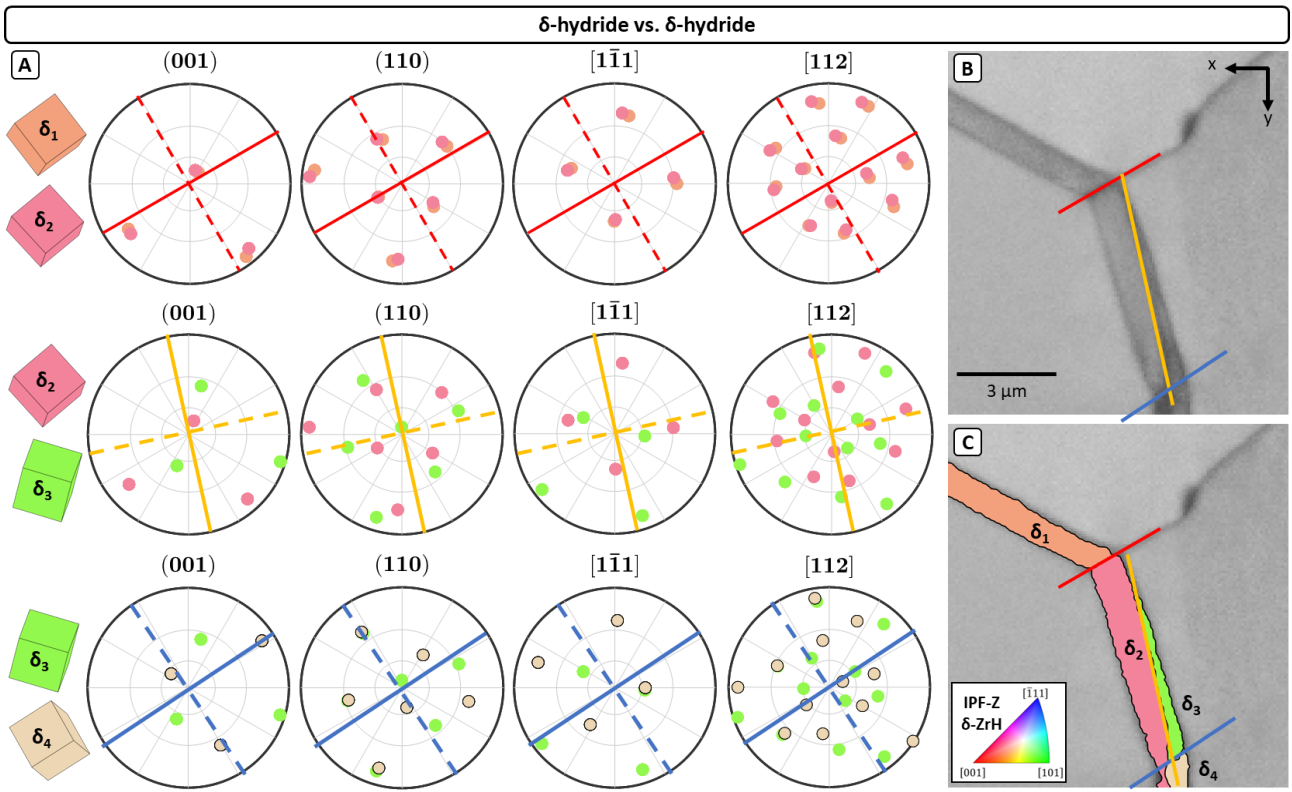

*Figure 7 - δ/δ orientation relationships in the pFIB hydride map. (A) pole figures for $\delta_1/\delta_2$, $\delta_2/\delta_3$ and $\delta_3/\delta_4$, with grain boundary traces overlaid and the perpendicular bisector (which is a geometrical construction used to reveal the zone of the interface plane). Traces are solid lines and the perpendicular bisectors are dashed lines of the same colour; (B) pattern quality with inter-hydride grain boundary traces identified; (C) pattern quality map overlaid with δ hydride grains, using IPF-Z colouring.*

HR-EBSD analysis is shown in Figure 8. The peak height map indicates that the hydride has poorer quality patterns than the matrix, and the righthand matrix grain has poorer quality patterns than the other two. The MAE map shows a significant concentration of deformation (indicated by poorer quality data) at the triple junction. All the other maps have been filtered to only show reasonable values, using a filter to remove points with a PH < 0.3, and a MAE > $5\times10^{-4}$.

The rotation tensor maps show that there is a lattice rotation associated with the hydride metal interface, and, to a lesser extent, the undecorated grain boundary. The stress field maps show a stress concentration at the top interface of the triple point, which is spatially localised to within 1.5 µm of the interface.

In the GND density map, the stress concentration has a corresponding peak region, but the hydrides also show higher (and fluctuating along the length) values than the surrounding matrix.

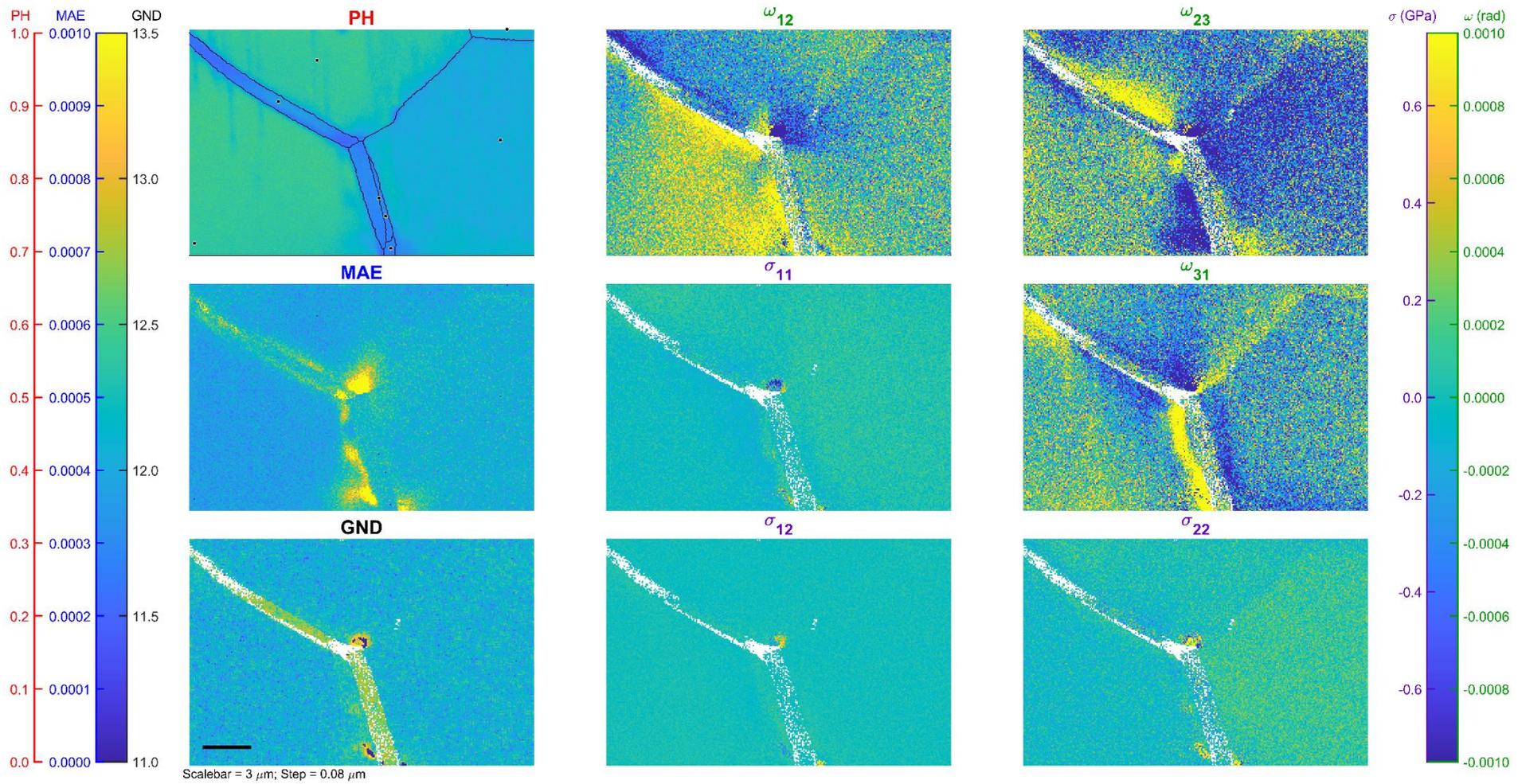

*Figure 8 - HR-EBSD analysis of triple junction hydride*

*. PH = peak height with reference points per mapped region plotted; MAE = mean angular error; GND = sum of GND density across all slip systems, reported as $\log_{10}$(dislocation length/volume, in per $m^2$); components of the rotation ($\omega_{ij}$, in radians) and in-plane stress tensor ($\sigma_{ij}$). Maps are plotted against different colour scales. The stress tensor and rotation tensor maps are shown with respect to the reference points, as indicated on the PH map.*

## 5.2 Hydrides that protrude into the Zr matrix

Two examples where the grain boundary hydride protrudes into the matrix: pinched off from smooth grain boundary hydrides, and hydride 'spikes' protruding from the grain boundary.

### 5.2.1 Example 3: (B2 – BIB polished): Pinched off hydride

In this map, there are two α grains with the grain boundary decorated by two hydride grains whose morphology aligns well with regards to the grain boundary until they 'pinch off' into the matrix at opposite ends of the hydride region (both deviations have the same angle into the matrix for this sectioned surface). On the righthand side of the mapped area, there is another region showing similar but less defined trends – the important point being that the hydride pinching off into the matrix is at the same angle and into the same grains. Note that the map is slightly spatially distorted due to drift during mapping – the insert at Figure 9 B shows the undistorted forescatter image of the hydride.

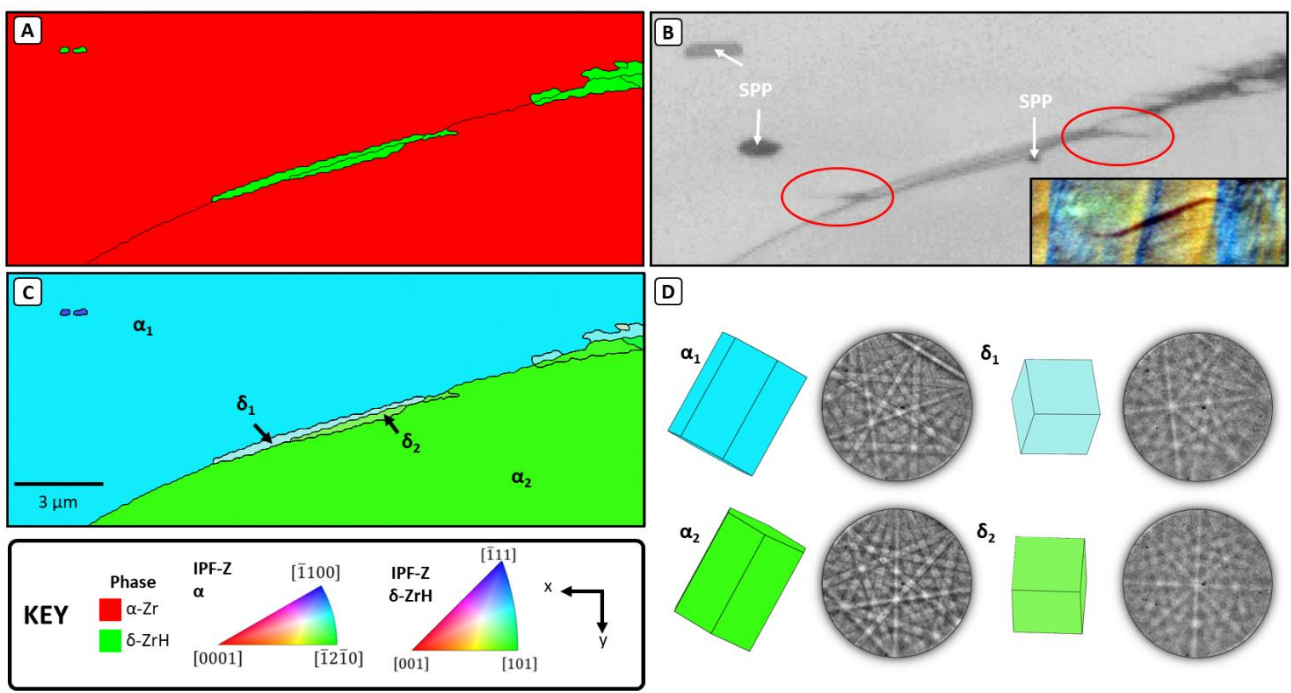

*Figure 9 - Orientation analysis of the pinched off hydride map.(A) Indexed phase map; (B) pattern quality map with SPPs indicated and red ovals highlighting the pinched off sections of hydride (insert: forescatter image of the hydride); (C) crystal orientation (IPFZ coloured) along with labels for key grains; (D) prisms and representative patterns for the identified grains.*

Figure 9 shows the orientations of the two adjacent α grains and the two main δ hydrides decorating this boundary. The pinched off sections are not visible on the orientation map (Figure 9 C) as they are too thin, but they are clearly identifiable on the pattern quality map due to the reduction in pattern quality compared to the matrix – they are also visible in the forescatter image. The orientations for the pinched off sections are assumed to be that of the hydride grain that they come from. The α orientations again appear close in orientation (see Figure 9 D) whilst the δ orientations also appear to be related, with a similar (001) plane orientation seen in the prisms.

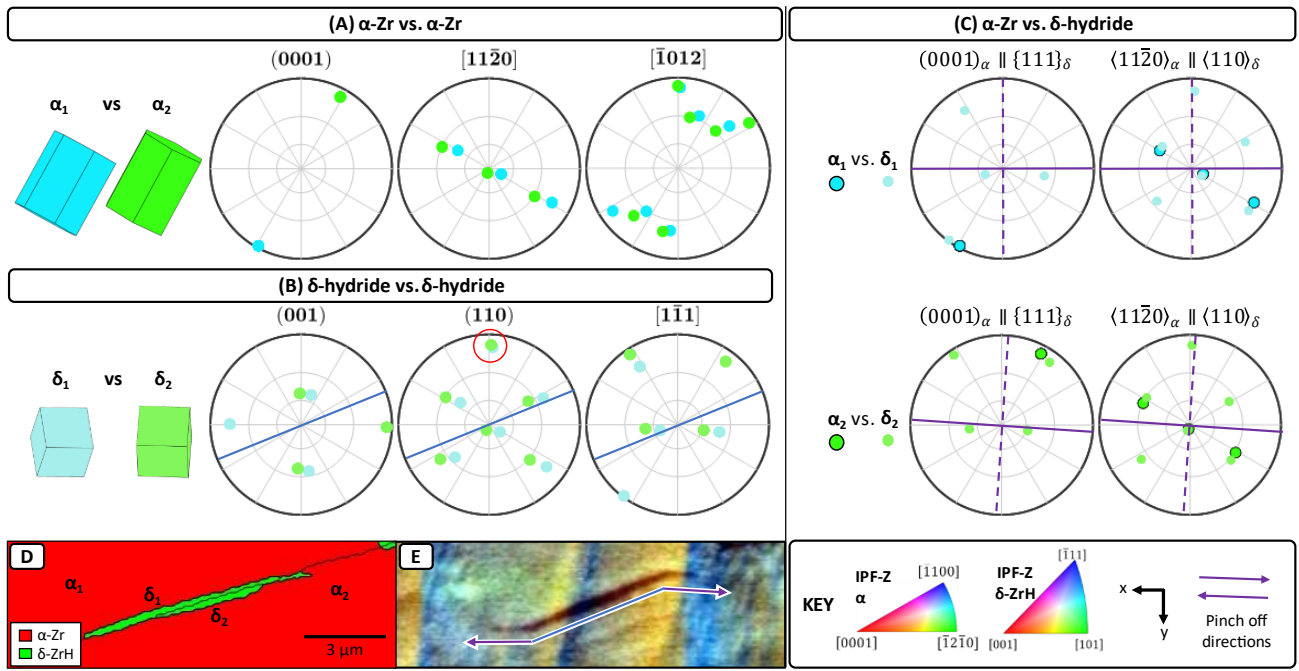

*Figure 10 - Orientation relationships in the pinched off hydride map. Pole figure analysis of (A) α-Zr/α-Zr OR;(B) hydride/hydride OR – with common point indicated by red circle and grain boundary trace overlaid; (C) α-Zr/δ-hydride OR – with pinched off direction (purple solid line) and zone of interest (dashed purple line) overlaid. (D) Phase map with grain labels; (E) FSD image with pinched off directions and δ/δ interface traces annotated.*

Analysis of the orientation relationships in this map are found in Figure 10. $\alpha_1/\alpha_2$ have a similar basal plane with a slight rotation between $\langle 11\bar{2}0\rangle_\alpha$ directions, similar to those seen in Figure 6 for the decorated α/α grain boundaries. $\delta_1/\delta_2$ have a shared $\{110\}_\delta$ plane and are rotated about this, by the same amount as $\alpha_1/\alpha_2$, suggesting that the two hydrides originate from different α grains (i.e. $\alpha_1$ and $\delta_1$ / $\alpha_2$ and $\delta_2$ ). In both cases, the hydride OR analysis in Figure 10 C fits well for $\langle 11\bar{2}0\rangle_\alpha \parallel \langle 110\rangle_\delta$ but not for $(0001)_\alpha \parallel \{111\}_\delta$. Future analysis will investigate alternative hydride/α-Zr ORs.

The pinched off hydride regions have a major axis that extends into the matrix, and these directions are indicated in Figure 10 D using purple arrows, and overlaid on the α vs δ pole figures in Figure 10 C, along with a perpendicular line indicating the zone of interest. This analysis indicates that both $\delta_1$ and $\delta_2$ have a $\langle 111\rangle_\delta$ direction along the zone of interest. It is also notable that these directions are very similar to an edge in the cubic crystal (i.e. [001] direction).

Figure 11 shows HR-EBSD analysis for the pinched off hydride map. The PH map again indicates that the hydride has poorer quality patterns than the matrix, as is the case for the SPP far away from the hydride. The pinched off hydride regions are also identifiable in the PH map, showing a lower value than the surrounding matrix. The MAE map shows concentrations between the pinched off hydride and the α/α grain boundary and extending from the tip of these points (and regions of high MAE can be related to large strain gradients within the interaction volume). All the other maps have been filtered to remove points with a PH < 0.3, and a MAE > $5\times10^{-4}$.

There are two regions of the hydride that are interesting in the HR-EBSD maps, (1) the two hydrides along the α grain boundary; (2) the two individual hydrides as they protrude into the grains.

For the hydrides along the α/α grain boundary, the rotation tensor maps show a similar trend to the previous examples, with a lattice rotation associated with the hydride metal interface. In this case this is more notable in the lower α grain, at the interface with $δ_2$. Here, the peak region magnitude reaches +/- $5×10^{-3}$ rad, and spatially extends ~1 μm from the boundary. Within the hydrides, the rotation tensor maps are relatively neutral along the length.

Along the smooth hydride region, the stress field maps show fluctuations of up to ~400 MPa with the highest values localised to within <0.5 μm of the α/δ interface, again to a slightly greater extent in the lower α grain. The stress fields are of opposite magnitude relative to their reference point in the α grains above $δ_1$ and below $δ_2$. GNDs are higher in the hydrides along this region compared to the matrix.

To explore the stress fields near the tips of the hydrides as the protrude into the grains, two zoomed areas (taken from the previous map shown in Figure 11) are reproduced in Figure 12 and Figure 13 to assist in the analysis of these subtle but exciting maps.

Figure 12 shows the HR-EBSD analysis for a zoomed in area on the lefthand hydride pinched off region (end of $δ_1$ in Figure 9). In this region, the hydride extends away from the grain boundary at an angle of approximately 20 ° (also 20 ° from the long axis of the hydride), and approximately along the $\langle 001 \rangle_δ$ crystallographic direction which is aligned along the x axis in Figure 12.

The PH and MAE maps in Figure 12 show that the data is reasonable for the matrix region near to the hydride, and also provide a visual indication of where the hydride tip is located.

The field map showing the density of GNDs is also spatially structured in the matrix, indicating that the GNDs may be related to the formation of low angle boundaries or slip activity on specific slip planes, related to the formation of the hydride and the shape & crystal orientation of the hydride and matrix phases. It is notable that these GNDs extend near the grain boundary, and away from the hydride tip, but they are not located in the matrix above the hydride-matrix interface ($α_1/δ_1$ – see Figure 9 for labelled figure).

The lattice rotation fields $(\omega_{ij})$ show heterogeneous lattice rotation associated with the tip of the hydride that extends horizontally away from the grain boundary and into the matrix phase. Note that these fields are measured relative to the reference point, and so variations (rather than the absolute values) in these maps are important. For the rotation fields, at the tip of the hydride there are two 'lobes' of +/- lattice rotation that extend +/- 30 ° and this is most easily seen in the $\omega_{23}$ field.

The spatial gradient of the lattice rotation field can be explained by (and is used to calculate the) storage of GNDs, and with a good model, these fields can be used to explore how much accommodation strain is required around an embedded hydride that forms in the solid state at elevated temperature.

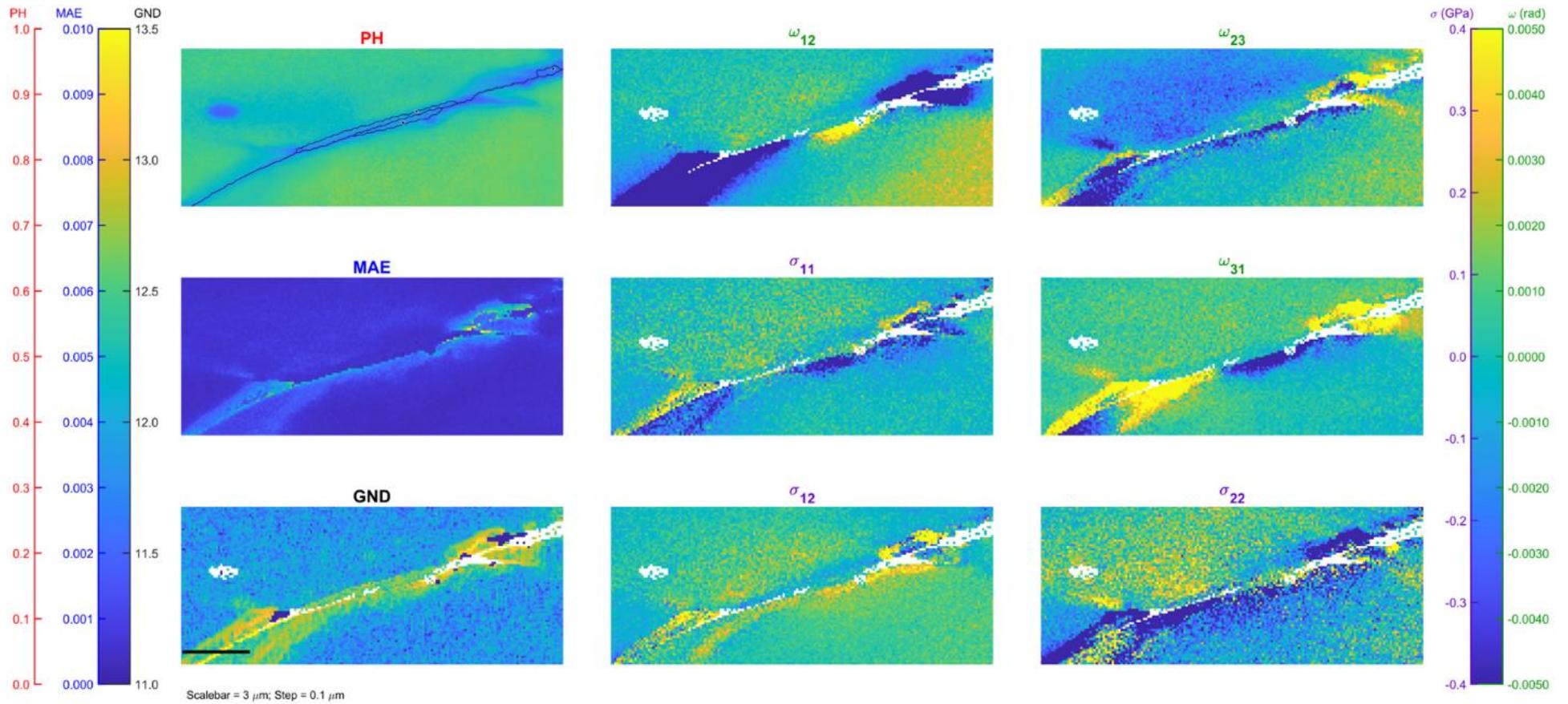

*Figure 11 - HR-EBSD analysis of the pinched off hydride map. PH = peak height with reference point*s per mapped region plotted; *MAE = mean angular error;* GND = sum of GND density across all slip systems, reported as log10(dislocation length/volume, in per m2); components of the rotation ($\omega_{ij}$, in radians) and in-plane stress tensor ($\sigma_{ij}$). Maps are plotted against different colour scales. The stress tensor and rotation tensor maps are shown with respect to the reference points, as indicated on the PH map.

Similar to maps of variations in lattice rotation, the stress fields $(\sigma_{ij})$ are measured with respect to the reference pattern and so only show the (in-plane) type 3 stress variations. These stress variations are asymmetric above and below the tip of the hydride, and their extent spatially aligns with the rotation fields. For example, it is noticeable that the $\sigma_{22}$ stress (i.e. the stress along the Y axis) in the upper lobe of the deformation field near the hydride tip is positive (i.e. relative tension) as compared to the lower lobe. Furthermore, the stress fields and rotation fields nearer the grain boundary are modulated strongly by the grain boundary and the presence of the hydride protrusion.

The residual stresses measured here are up to +/- 400 MPa which may be considered high for Zircaloy-4 (yield stress = between 241 and 455 MPa [32], [33]) but note that locally higher stresses may be found if there is residual work hardening (i.e. dislocation structures that increase the stored stress state prior to further plastic strain) and that zirconium has anisotropic plastic deformation, which can result in locally higher stresses (as plastic deformation can only relieve the shear stress on a specific slip plane/combination of shear stresses).

These lobes of high stress, GND activity, and lattice rotation typically extend ~1.5 µm away from the hydride tip. The distance indicated here will be slightly affected by the resolution of the method (i.e. a stress applied to an elastic solid will asymptotically approach zero far field stress) and the choice of colour scale.

Figure 13 contains the area around the other pinched off hydride tip, this time for α$_2$/δ$_2$ (see Figure 9). The two hydrides (δ$_1$ and δ$_2$) have grown sympathetically along the grain boundary and pinched off at opposite ends, at approximately the same angle into the matrix in both cases, i.e. along the x axis. This leads to some interesting similarities and differences. The maps are spatially similar, although flipped (hydride points to the right instead of the left), with similar length into the matrix and angle relative to the grain boundary. The rotation and stress fields show the +/- lobes at the hydride tip and are of similar magnitude.

The main difference is that the +/- regions are opposite in both $\omega_{ij}$ and $\sigma_{ij}$. For example, $\omega_{31}$ in Figure 13 has a positive value between the hydride and α/α grain boundary, with negative values on the α$_2$ matrix side of the hydride, whereas $\omega_{31}$ in Figure 12 has the opposite. These differences are seen because the hydride that forms results in a different accommodation of the misfit strain in the α-Zr matrix. This is likely due to the difference in α grain orientation – there is a ~7 ° difference in $(0001)_\alpha$ with ~16 ° minimum angle between $\langle 11\bar{2}0 \rangle_\alpha$ directions between α$_1$ and α$_2$ (see Figure 10).

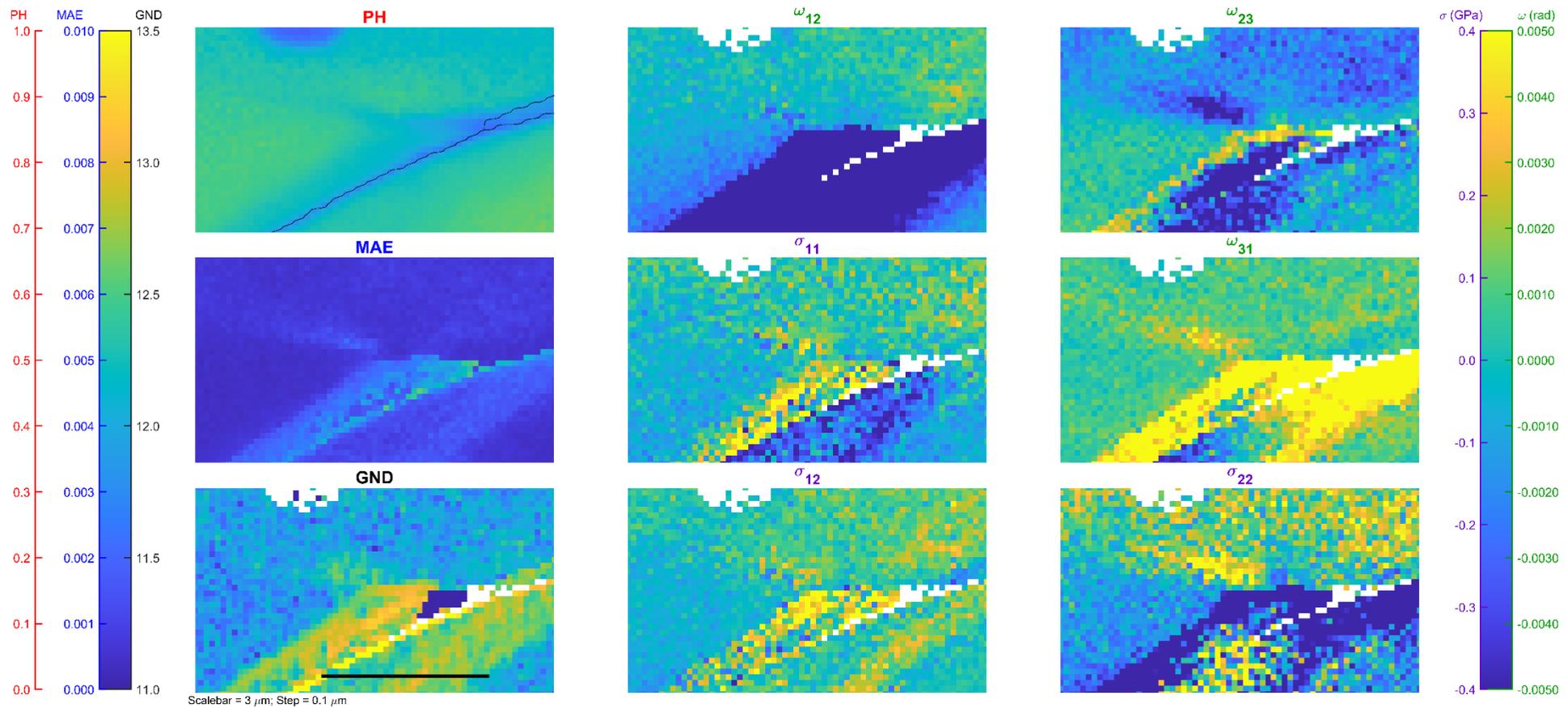

*Figure 12 - HR-EBSD analysis of the pinched off hydride map – Tip 1*

*. PH = peak height with reference points per mapped region plotted GND = sum of GND density across all slip systems, reported as $\log_{10}$(dislocation length/volume, in per $m^2$); components of the rotation ($\omega_{ij}$, in radians) and in-plane stress tensor ($\sigma_{ij}$). Maps are plotted against different colour scales. The stress tensor and rotation tensor maps are shown with respect to the reference points, as indicated on the PH map in Figure 11.*

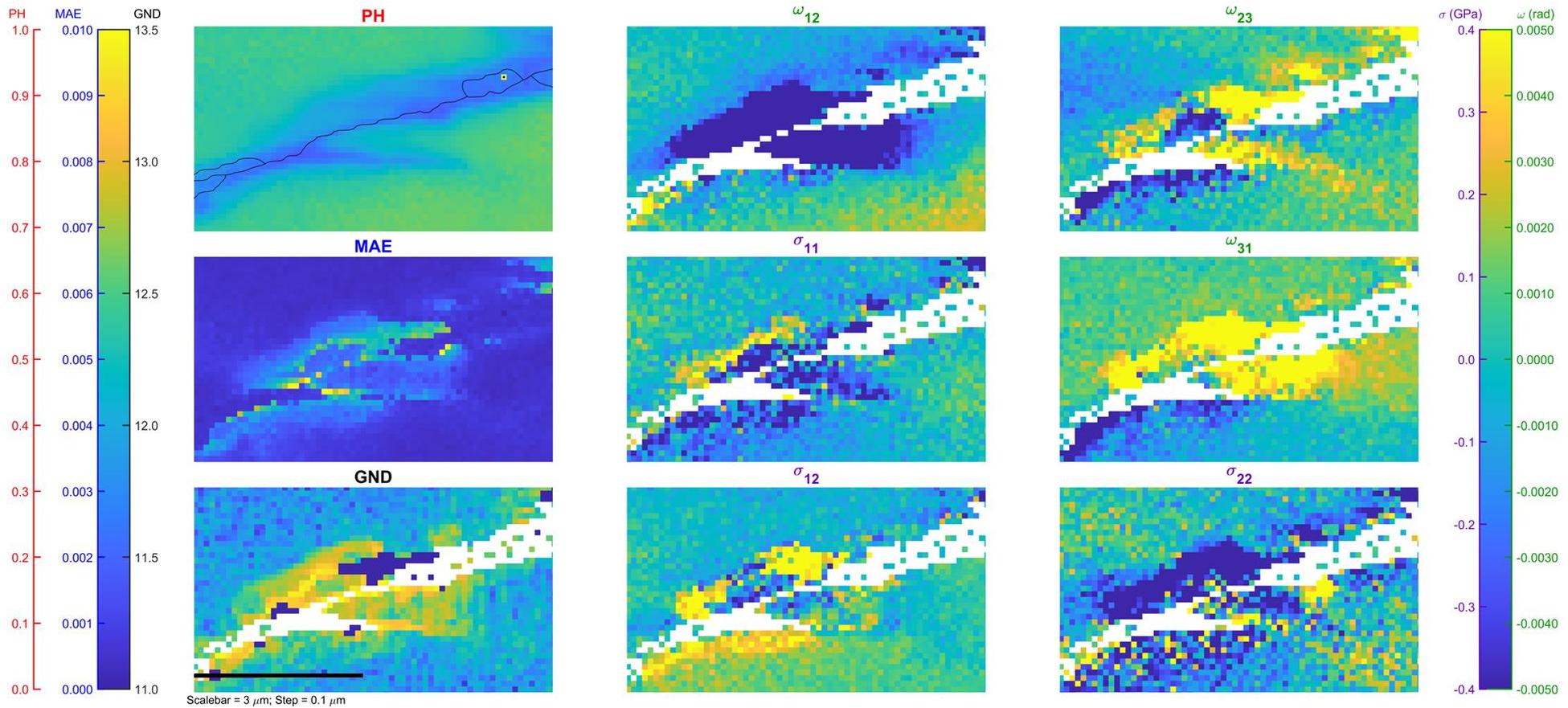

*Figure 13 - HR-EBSD analysis of the pinched off hydride map – Tip 2*

*. PH = peak height with reference points per mapped region plotted; GND = sum of GND density across all slip systems, reported as $\log_{10}$(dislocation length/volume, in per $m^2$); components of the rotation ($\omega_{ij}$, in radians) and in-plane stress tensor ($\sigma_{ij}$). Maps are plotted against different colour scales. The stress tensor and rotation tensor maps are shown with respect to the reference points, as indicated on the PH map in Figure 11.*

### 5.2.2 Example 4: (B3 – BIB polished) 'Spiky' hydride

The final example is another region where there are 'thin spikes' of hydride starting from the grain boundary and going into one of the two α grains, as shown in Figure 14 and called the 'spiky hydride' example. These hydrides do not show very well in the processed grain map (Figure 14 A & C) due to their size and aspect ratio, but they are visible in the pattern quality map (Figure 14 B). The hydride spikes protrude up to 13 µm into the matrix in this map at an (in plane) angle of ~40 ° to the α/α grain boundary.

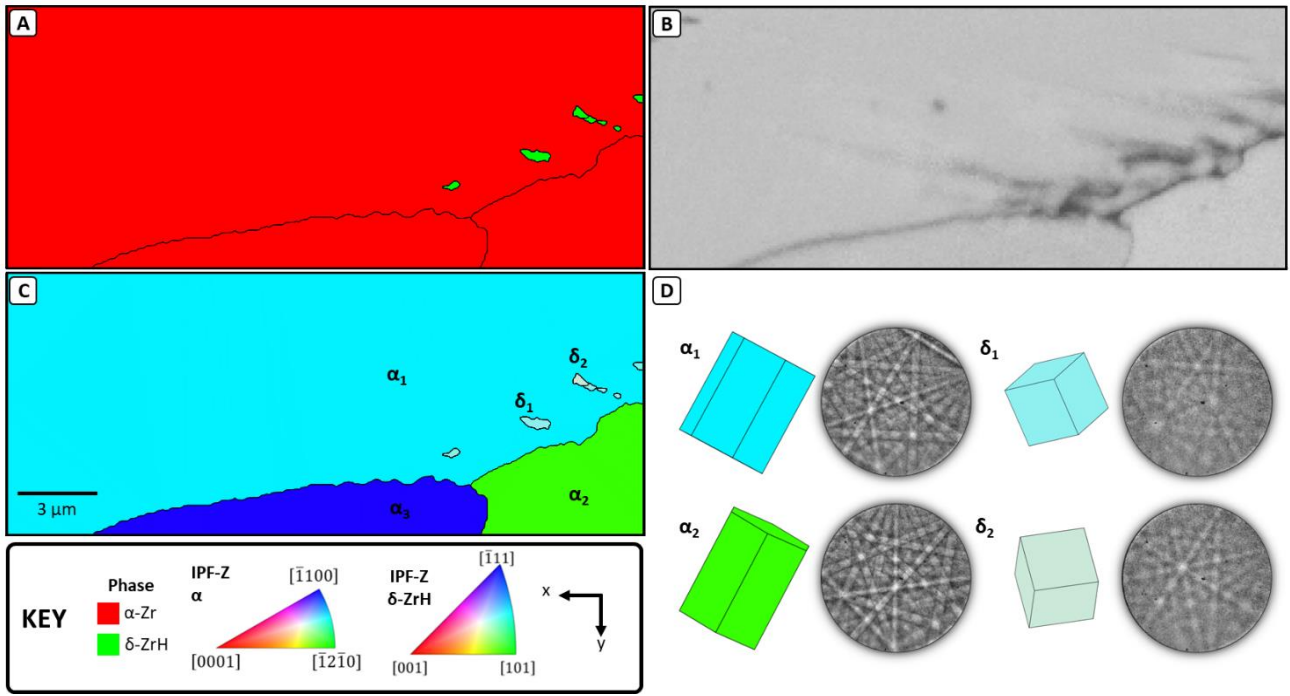

*Figure 14 - Orientation analysis of the spiky hydride map. (A) Indexed phase map; (B) pattern quality map; (C) crystal orientation (IPFZ coloured) along with labels for key grains; (D) prisms and representative patterns for the identified grains.*

Figure 14 contains orientation analysis for the two main α grains at this point, along with the two hydride orientations that index (in the widest sections of the hydrides). $α_1$ and $α_2$ have similar orientations, with a slight rotation (6 °) about $\langle 0001 \rangle_α$ direction and similar basal planes. The δ hydrides appear to share a [111] direction about which they are rotated.

Analysis of the orientation relationships between the δ hydrides, the α grains, and the α/δ interfaces are shown in Figure 15.

The α grain pair have a common $\{10\bar{1}2\}_α$ plane (2.7 ° angle between) and a similar basal plane direction (6.5 ° offset), rotated about this axis to give a 16.0 ° angle between the <a> directions. The δ grains (part of the same spikes) are twins, with a shared $\langle 1\bar{1}1 \rangle_δ$ direction and three shared $\{1\bar{1}0\}_δ$ planes.

Analysis of the α/δ OR (Figure 15 C) shows that both hydride phases fit very well with the hydride OR with $α_1$ which is expected considering both types of hydride protrude into $α_1$.

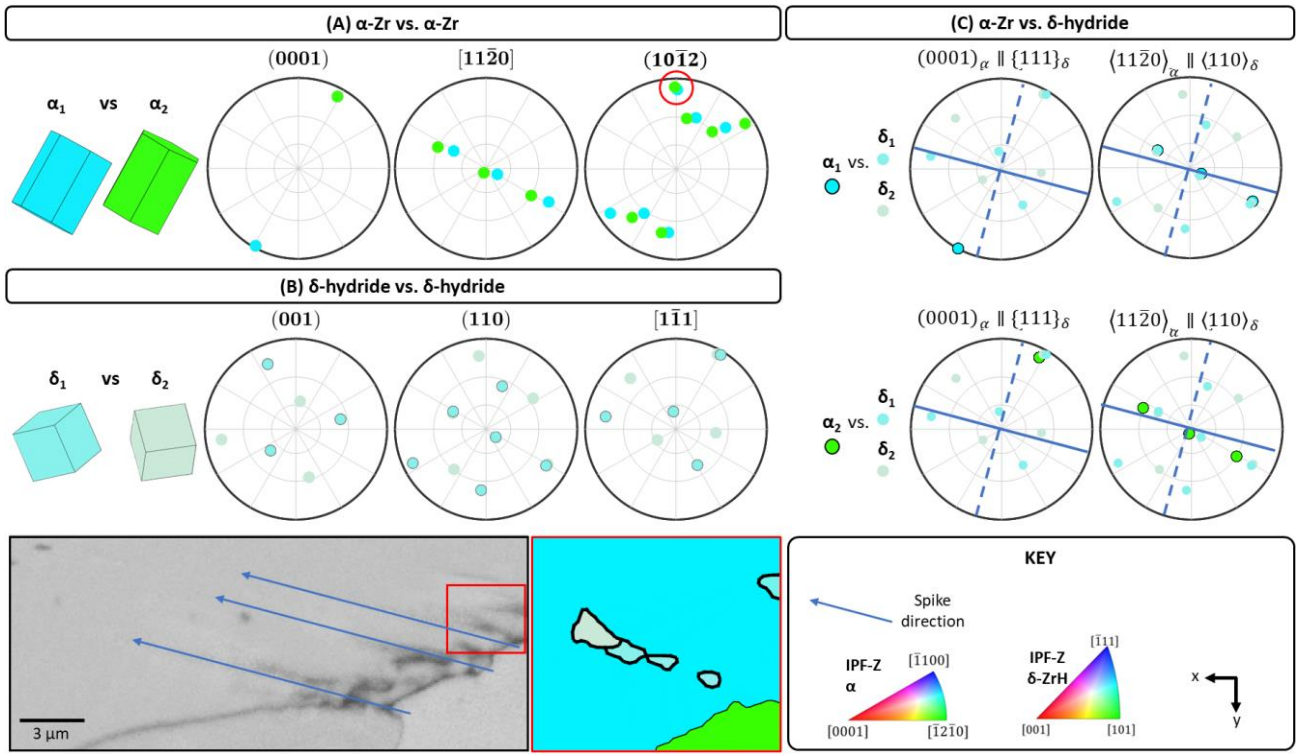

*Figure 15 - Orientation relationships in the spiky hydride map. Pole figure analysis of (A) α-Zr/α-Zr OR;(B) hydride/hydride OR; (C) α-Zr/δ-hydride OR. (D) Pattern quality map with hydride directions annotated; (E) Zoomed in orientation map (IPFZ colouring) showing the two hydride orientations within a hydride spike.*

HR-EBSD for the spiky hydride map is shown in Figure 16, where PH and MAE show that the data is reasonable. The PH map reveals the hydride locations, through the poorer quality patterns for the hydrides. All the other maps have been filtered to remove points with a PH < 0.3, and a MAE > 5x10$^{-4}$.

Although there are multiple hydrides along this section of grain boundary which leads to come potential interactions between them, the general trends seen are similar to the previous example (Figure 11). Specifically, for the better defined hydrides, they resemble tip 2 in Figure 13 with similar trends in the lattice rotation maps $(\omega_{ij})$ for the relative +/- regions. For the stress fields $(\sigma_{ij})$, again similarities to Figure 13 are seen, with negative stress values (relative to the reference point) observed between the hydride and the α/α grain boundary. These residual stress values are again seen locally to be +/- 400 MPa.

The biggest difference is the missing 'lobes' of +/- lattice rotation and stress in the maps in Figure 16. Although they are not immediately apparent, the zoomed in maps for one of the tips, shown in Figure 17, suggests that they are present, but much smaller in size and magnitude (see $\sigma_{12}$ for example). This difference could be due to the relative hydride size, interaction of the multiple hydrides in the region, or the crystallographic orientations involved.

The map of stored GND density indicates that regions of high GND density are near the tips of the hydrides and tend to extend ahead of each of the hydride tips. These GNDs seem to be confined to one crystallographic direction (or plane) which could be consistent with accommodation of the misfit strain ahead of the hydride.

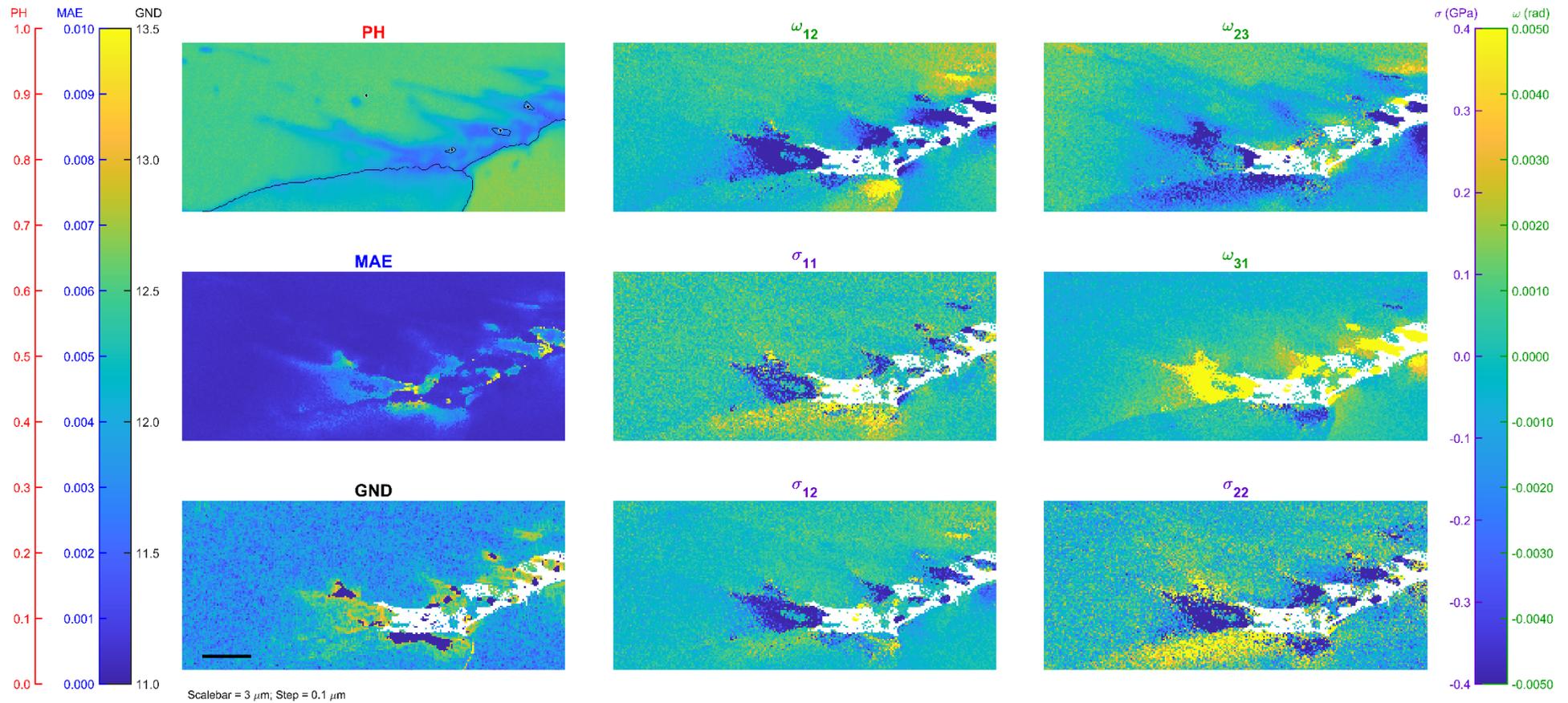

*Figure 16 - HR-EBSD analysis of the spiky hydride map*

*. PH = peak height with reference points per mapped region plotted; MAE = mean angular error; GND = sum of GND density across all slip systems, reported as $\log_{10}$(dislocation length/volume, in per $m^2$); components of the rotation ($\omega_{ij}$, in radians) and in-plane stress tensor ($\sigma_{ij}$). Maps are plotted against different colour scales. The stress tensor and rotation tensor maps are shown with respect to the reference points, as indicated on the PH map.*

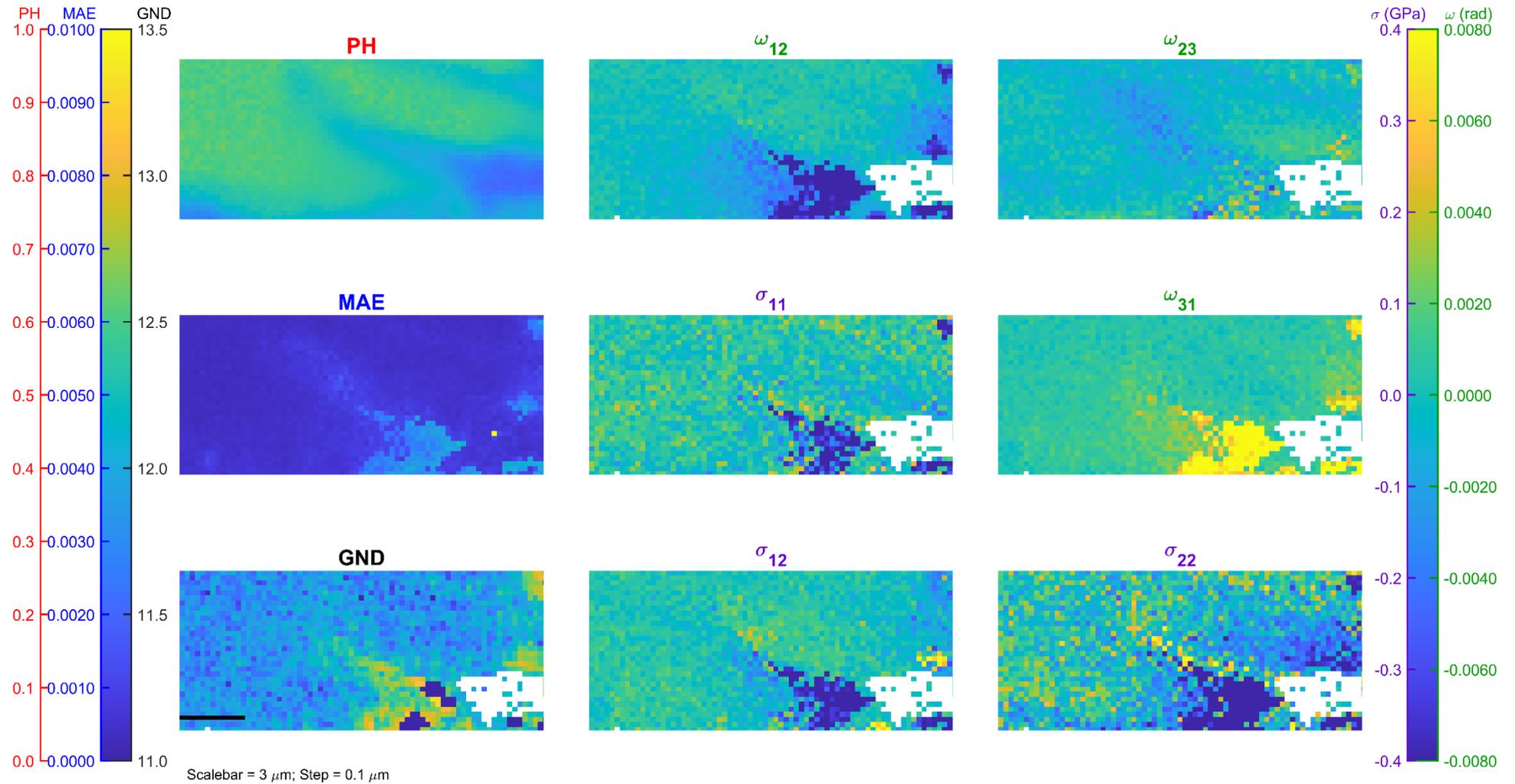

Figure 17 - HR-EBSD analysis of a section of the spiky hydride map. PH = peak height with reference points per mapped region plotted; MAE = mean angular error; GND = sum of GND density across all slip systems, reported as $\log_{10}$(dislocation length/volume, in per $m^2$); components of the rotation ($\omega_{ij}$, in radians) and in-plane stress tensor ($\sigma_{ij}$). Maps are plotted against different colour scales. The stress tensor and rotation tensor maps are shown with respect to the reference points seen in Figure 16.

# 6   Discussion

This paper shows, via electron backscatter diffraction measurements, that accommodation of δ-hydrides within Zircaloy-4 can be associated with and without local variations in residual stress and stored GND content. In the present work, all these hydrides are considered 'macroscopic' as they are observed using an EBSD step size of 80-100 nm. These observations indicate that the successful growth of large δ-hydrides within the zirconium matrix is influenced by the ability for the local microstructure to accommodate the misfit strain. In these cases, for hydrides at and near grain boundaries, this misfit strain seems to influence the shape of the grown hydride, and furthermore it may influence the initiation and propagation of microstructurally sensitive cracks [34].

These measurements are only achieved due to the quality of the sample preparation, and equal polishing of the hydride and metal using cryogenic temperatures. This is achieved by cross sectioning using broad and plasma-based focussed (Ar) ion beam polishing, with the sample held at cryogenic temperatures.

The two methods provide different and complimentary opportunities to enhance our understanding of materials:

BIB produces high quality surfaces over large areas which can be polished at the same time, but these surfaces are not created with in situ monitoring and so finding a recipe that works well and targets specific regions is challenging. BIB also requires larger pieces and mounting/gluing, which may not be appropriate if used on active materials if you want to reduce sputtered material and reduce analysed volumes to be transferred into another instrument.

pFIB can provide access to 100s of μm of polished surface, with typically a shallow polishing depth, and this reduces the amount of material that can be studied from one polished surface. However, you can monitor polishing online during the creation of the surface which is useful when you need to polish to obtain the best quality data for a specific area of interest. As one example, this can be useful to target triple junctions for example. Furthermore, there are cases where the reduced volume of sputtered material could be advantageous, for example for preparing smaller volumes of material where the radioactivity of the sample is important.

For the present work, these two methods are exciting as they have enabled us to more systematically study hydride-metal interfaces, e.g. as a function of grain boundary misorientation and grain-boundary plane similar to prior studies of stress fields that extend from a blocked slipped band [35]. For the fuel cladding community, measurement of these stress-fields (especially for samples where reorientation has been observed) could be useful in providing stronger mechanistic insight into the reorientation mechanism, and supplement arguments regarding the interfacial energy [12].

The distribution and magnitude of these strains is strongly correlated with the location of the hydrides within the microstructure, and the morphology of the hydride/zirconium interfaces and the associated orientation relationships.

For the first hydride decorating along a single grain boundary (Figure 3) the grain boundary pair has a similar <c> axis alignment either side of the boundary and the hydride shows an orientation relationship which is well aligned with one of the grains ($\alpha_1$). The interface of the hydride and parent grain ($\alpha_1$-$\delta_2$) is flat and likely contains the habit plane of the hydride (the perpendicular bisector of the hydride habit plane and OR is consistent). This well-matched alignment and interface is likely why the magnitude of the accommodation stress is lower on this interface, and lower than the other hydride-matrix interfaces from the other mapped regions. This fits with analysis that links the coherency of the hydride-matrix interface with strain, for example, Barrow et al. [12] show an example where there is a fully coherent interface between $(11\bar{2}0)_\alpha \parallel (2\bar{2}0)_\delta$.

If the hydride is well accommodated with the local orientation relationship, especially if this is accommodated well by neighbour grains at a grain boundary, then the hydride seems to decorate the grain boundary well with limited and smoothly varying strain heterogeneity extending into the matrix for the case where the hydrides decorate the triple junction interface, and the hydride variants all can be related to the matrix orientations (see Figures 5-8). The termination of these hydrides and the meeting of the hydrides at the matrix-triple junction does result in a greater amount of strain heterogeneity, and the storage of geometrically necessary dislocations (see Figure 8) which is consistent with misfit of multiple hydrides at the junction.

When the hydride starts to protrude into the matrix, there is a change in both the local orientation relationship and the strains around the hydride. A nice example of this is the pair of hydrides that are smoothly decorating then pinch off at the ends (example 3), protruding into the adjacent grain (from which each originated). The smoothly decorating section shows similar trends to previous, but there is significant strain heterogeneity where the hydride protrudes into the matrix. At the hydride tips, there are two 'lobes' of +/- high stress, GND activity, and lattice rotation extending away from the hydride tips. These lobes are consistent with literature examples of lenticular hydrides within the $\alpha$-Zr matrix, where similar is seen at the hydride tips revealed with complementary analysis [12], [15].

For example 3, the two hydrides enable us to explore the relationship between the OR and the variation in lattice rotation and residual stress. Here, the two hydrides are well aligned, sharing a (110) plane, and the OR between $\delta_1$ and $\alpha_1$ fits well with the hydride OR. However, for the lower hydride ($\delta_2$) there is a deviation from the hydride OR. Looking closer at Figure 9, there is a ~5° rotation of $\delta_2$ along its length, with the highest deviation from the hydride OR at its centre, and this reduces as the hydride protrudes into the matrix and away from the grain boundary. The morphology and OR of this hydride fits well with the stress field seen (in Figure 11), with higher stress at the centre where there is most deviation from the hydride OR, and lower stress towards the tip as it rotates (before the hydride tip effects occur). This means that the grain boundary and orientations either side therefore have an impact on whether the hydride smoothly decorates or protrudes into the matrix, and the local strains seen.

In contrast, example 4 contains a very similar grain pair, yet we only observe hydride protruding into the matrix in the equivalent of $\alpha_1$ (see Figure 14). The reason for this difference is likely due to a small variation in grain orientation, however, we can see by comparing examples 3 and 4 that it is less favourable to precipitate and grow a hydride in the adjacent grain unless it has already grown from another hydride for this grain orientation pair.

Further grain orientation effects can be seen by comparing the stress and rotation fields around the hydride tips between the two hydrides that have grown cooperatively along the grain boundary in example 3 (Figures 12 and 13). In this case, despite the regions around the two hydride tips being of similar magnitude, the +/- regions are opposite in both $\omega_{ij}$ and $\sigma_{ij}$, likely due to the ~7 ° difference in $(0001)_\alpha$ with ~16 ° minimum angle between $\langle 11\bar{2}0 \rangle_\alpha$ directions between α$_1$ and α$_2$ (see Figure 10).

For hydride precipitation, in the literature is often commented that as hydrides precipitate, they can accommodate significant misfit strain (e.g. [8], [12], [15], [36]) through the nucleation of dislocations. Similar to this hypothesis, we see significant lattice rotation gradients which are likely supported by GNDs (and we report them as GNDs, as per Nye's analysis). Importantly, HR-EBSD cannot easily analysis the presence of SSDs, e.g. small dislocation loops, that may be below the step size of the EBSD measurement grid and so there may be some dislocations that are missed in our analysis.

## 7   Conclusion

In-service, hydrogen ingress into Zr alloys can lead to the formation of brittle hydride phases which have a deleterious effect on the component performance. The misfit strain between hydrides and the matrix is thought to cause a localised stress field, which can influence hydride nucleation, growth, and reorientation. In this work, conventional EBSD and HR-EBSD were used to characterise hydride-matrix deformation fields near grain boundaries. These fields vary for hydrides that smoothly decorate the grain boundary and those that protrude into the matrix, providing insight into hydride-microstructure influenced component performance.

- For smoothly decorating grain boundary hydride, HR-EBSD analysis showed very small variations in stress, lattice rotation and GND density associated with the (α/δ) grain boundary interfaces. These hydrides were much larger than previously analysed hydrides, however the behaviour is not unexpected due to both the misfit strain of the hydride (causing the stress) and the morphology of the hydride (it smoothly decorates the grain boundary).
- For hydride protruding into the matrix, there are large stresses and stored GND density that extend ahead of the tips of the hydride. These results show similarity to the experimental [12] and modelling [15] work on small lenticular hydrides within the α matrix.

For more complex hydride/matrix structures, and the deformation fields are consistent ahead of hydride tips of the same hydride type. These results may enable enhanced understanding of the role of hydrides in fracture as well as stress-induced hydride reorientation, as likely found in delayed hydride cracking.

## 8  Acknowledgements

We acknowledge funding and support from Rolls-Royce plc. We have had helpful and insightful discussions with Mike Rogers and Prof Fionn Dunne on this work. We gratefully acknowledge the Engineering and Physical Science Research Council for funding the Imperial Centre for Cryo Microscopy of Materials at Imperial College London (EP/V007661/1) and for funding the MIDAS program grant (EP/S01702X/1).

## 9  Author Contributions

Ruth Birch – conceptualisation, formal analysis, investigation, methodology, visualization, writing – original draft, review & editing; James Douglas – investigation, methodology, writing – review & editing; Ben Britton – conceptualization, software, supervision, writing –review & editing.

**Supplementary material 1**

Table 5 contains the sample preparation details for samples A and B. Exact hydrogen content was not checked, but is approximated based on previous work [2].

*Table 5 – Sample preparation details for sample A and B*

| Sample | Microstructure (start) | Sample dimensions L x W x D | Heat Treatment(s) | Hydriding Details | Final polishing step |
|---|---|---|---|---|---|
| A | Fine grain α | 4 x 3 x 1.5 mm | Blocky-α HT: 2 weeks @ 800 °C / air cool  Hydride anneal: 7 hrs @ 475 °C  Cooling rate: slow furnace cool | No of sides: 1  Surface area hydrided: 12 mm$^2$  Time: 24 hrs  Expected [H] *: ~120 wt-ppm | pFIB |
| B | Fine grain α | 20 x 3 x 3 mm | Blocky-α HT: 2 weeks @ 800 °C / air cool  Hydride anneal: 3 hrs @ 460 °C  Cooling rate: 5°C/min (furnace cool) | No of sides: 2  Surface area hydrided: 120 mm$^2$  Current: 240 mA  Time: 24 hrs  Expected [H]*: ~240 wt-ppm | BIB |

* see [2] for DSC results used to approximate hydrogen content.

**Supplementary material 2**

*pFIB vs. BIB final polishing*

For this work, two argon ion beam polishing methods were used to prepare the final surface of the samples for EBSD analysis: pFIB and cross-section BIB. The final step for both methods was carried out at cryo temperatures to reduce ion-surface damage and secondary formation of hydrides (see [20], [21]).

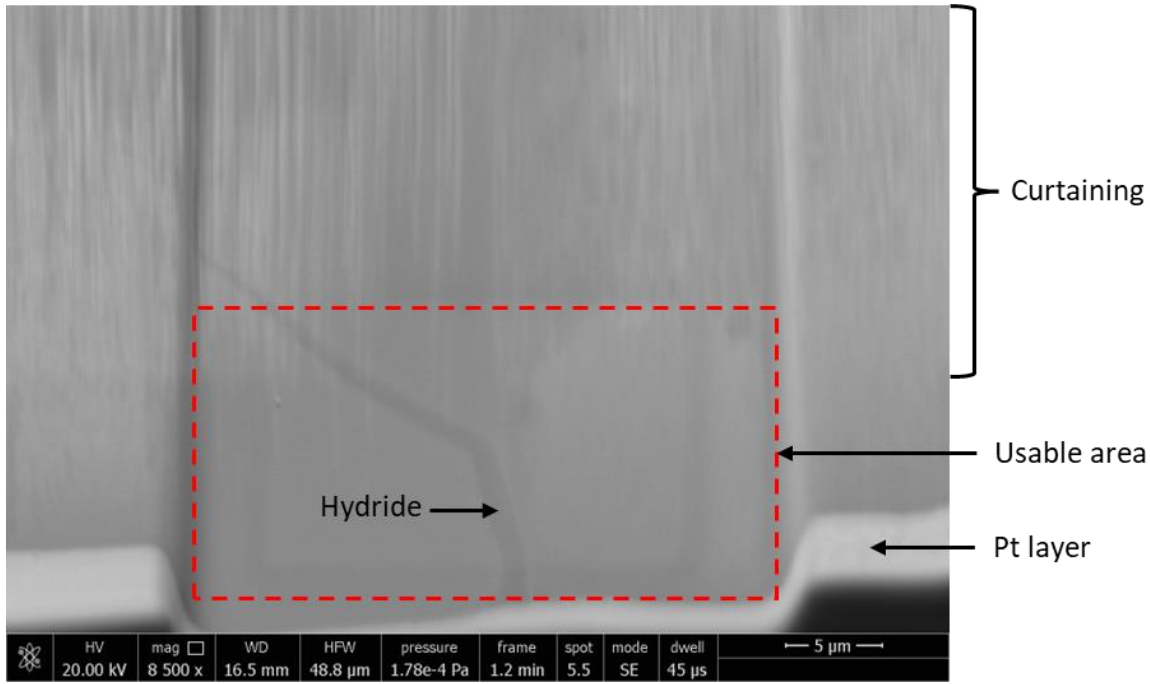

*Figure 18 - pFIB hydride area overview. SE image showing the polished hydride region, with the useable area indicated by the red dashed box and the start of significant curtaining indicated. Scalebar = 5 µm.*

The pFIB polishing (see Figure 18) provided a polished region of approx. 30 x 15 µm$^2$ (red dashed area in Figure 18). This region provided high quality patterns and indexed very well within this region (up to 100 % indexing success). Outside this area, surface roughening when cross sectioning (so called 'waterfall' effects or 'curtaining') caused significant local surface roughness. Rocking milling has been shown to reduce effects of curtaining in cross sections but this typically requires the use of a specialist stage which is not compatible with the cryo stage.

The BIB polished sample (Figure 19) provided a larger area, but had a rougher surface finish than the high quality area prepared using the pFIB. The BIB polished (semi-circular) region had maximum dimensions of 2.3 x 0.6 mm, and indexing was >90%. This large area provided multiple good regions for mapping, but also contained some ridges (i.e. significant local roughness) that meant some areas were not suitable for HR-EBSD analysis.

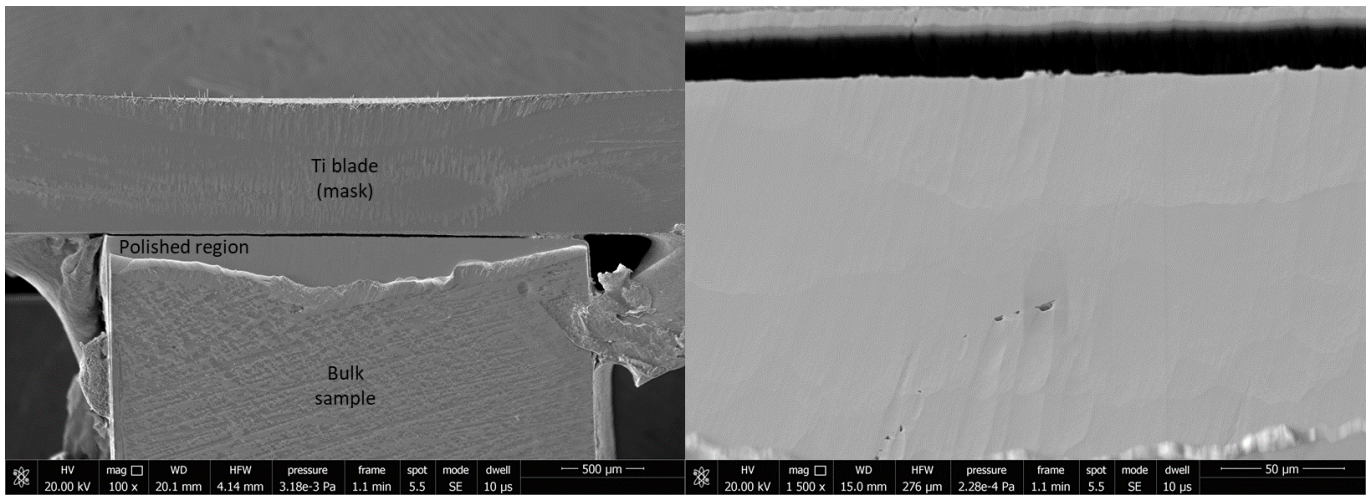

*Figure 19 – BIB cross-sectioned surface. (A) Secondary electron (SE) image showing the location of polished region whilst sample is attached to Ti blade that acts as a mask during cross-sectioning. (B) higher magnification SE image of the polished region showing some localised surface roughness.*